\documentclass[12pt]{article}

\usepackage{newtxtext,newtxmath}
\usepackage{graphicx}
\usepackage{caption}

\usepackage{soul}
\usepackage[letterpaper,margin=1in]{geometry}

\renewenvironment{abstract}
	{\quotation}
	{\endquotation}

\date{}

\usepackage[numbers,super,sort&compress]{natbib}

\usepackage{url}

\def\nattitle{
	Super-resolution femtosecond electron diffraction reveals electronic and nuclear dynamics at conical intersections
}
\title{\bfseries \boldmath \nattitle}

\author{
	Hui Jiang$^{1,2,3\dagger}$,
	Juanjuan Zhang$^{4\dagger}$,
    Tianyu Wang$^{2,3\dagger}$,
    Jiawei Peng$^{4}$,
    Cheng Jin$^{2,3}$,\and
    Xiao zou$^{2,3}$,
    Pengfei Zhu$^{1,2}$,
    Tao Jiang$^{2,3}$,
    Zhenggang Lan$^{4\ast}$,
    Haiwang Yong$^{5\ast}$,\and
    Feng He$^{2\ast}$,
	Dao Xiang$^{1,2,3\ast}$\and
    \small$^{1}$Tsung-Dao Lee Institute, Shanghai Jiao Tong University, Shanghai 201210, China.\and
    \small$^{2}$Key Laboratory for Laser Plasmas (Ministry of Education) and School of Physics and Astronomy,\and
    \small Collaborative innovation center for IFSA (CICIFSA),\and
    \small Shanghai Jiao Tong University, Shanghai 200240, China.\and
    \small$^{3}$Zhangjiang Institute for Advanced Study, Shanghai Jiao Tong University, Shanghai 201210, China.\and
    \small$^{4}$SCNU Environmental Research Institute,\and
    \small Guangdong Provincial Key Laboratory of Chemical Pollution and Environmental Safety, \and 
    \small MOE Key Laboratory of Environmental Theoretical Chemistry,\and
    \small School of Environment, South China Normal University, Guangzhou 510006, China.\and
    \small$^{5}$Department of Chemistry and Biochemistry, University of California San Diego, La Jolla, CA 92093, USA.\and
    \small$^\ast$Corresponding authors: zhenggang.lan@m.scnu.edu.cn (Zhenggang Lan);\and
    \small hyong@ucsd.edu (Haiwang Yong); fhe@sjtu.edu.cn (Feng He); dxiang@sjtu.edu.cn (Dao Xiang)\and
	\small$^\dagger$These authors contributed equally to this work.
}

\begin{document} 
%\linenumbers

\maketitle

\clearpage
\begin{abstract} %\bfseries \boldmath
Conical intersections play a pivotal role in excited-state quantum dynamics. Capturing transient molecular structures near conical intersections remains challenging due to the rapid timescales and subtle structural changes involved. We overcome this by combining the enhanced temporal resolution of mega-electron-volt ultrafast electron diffraction with a super-resolution real-space inversion algorithm, enabling visualization of nuclear and electronic motions at conical intersections with sub-angstrom resolution, surpassing the diffraction limit. We apply this technique to the textbook example of the ring-opening reaction of 1,3-cyclohexadiene, which proceeds through two conical intersections within 100 femtoseconds. The super-resolved transient structures near conical intersections reveal a C-C bond length difference of less than 0.4$~\text{\AA}$ and an approximately 30-femtosecond traversal time of the nuclear wave packet between them. These findings establish super-resolution ultrafast scattering as a transformative tool for uncovering quantum dynamics in molecules and open new avenues for studying light-matter interactions at the most fundamental level.
\end{abstract}

%%%%%%%%%%%%% MAIN TEXT %%%%%%%%%%%%%%%%%%%%%%%%%%%%%%
%\section*{Main}

\section*{Introduction}

\noindent
Conical intersections (CIs) are seams of degeneracy where two or more potential energy surfaces (PESs), representing different electronic states, intersect in molecules. They are characterized by strong non-adiabatic couplings between electronic and nuclear motions, enabling efficient energy conversion between these degrees of freedom~\cite{Bernardi1996}. Such intersections often facilitate ultrafast non-radiative decay, which is accompanied by bond breaking or isomerization processes~\cite{Levine2007}.These processes also play a critical role in photoprotection mechanisms in biomolecules~\cite{Crespo2004} and primary events in human vision~\cite{Polli2010}, making CI dynamics a central focus of research in photophysics, photochemistry and photobiology~\cite{Yarkony1996,Worth2004,Schuurman2018}.
Due to the nonadiabatic coupling extending beyond the Born-Oppenheimer approximation, it is imperative to simultaneously capture electronic and structural dynamics to gain a comprehensive understanding of CI dynamics. Despite considerable recent advances in time-resolved spectroscopic~\cite{Zinchenko2021,Worner2011} and diffraction experiments~\cite{Minitti2015,Yang2018,Hosseinizadeh2021}, the real-space imaging of ultrafast electronic and nuclear motions at CIs, with femtosecond temporal resolution and atomic spatial resolution, remains a significant challenge.

The photoinduced electrocyclic reaction of 1,3-cyclohexadiene (CHD) yielding 1,3,5-hexatriene (HT) has long been considered a textbook example for studying nonadiabatic pericyclic reactions~\cite{Hoffmann1968,Weber2011}. This provides significant insights into the mechanisms underlying vitamin D synthesis~\cite{Havinga1961}, organic synthesis~\cite{Chopade2006}, and the functioning of a wide range of molecular devices~\cite{Irie2014}.
Extensive studies have been conducted to elucidate the ultrafast CI dynamics involved in the ring-opening of CHD, with the established reaction pathway~\cite{Garavelli2001,Weber2011} schematically depicted in Fig.~\ref{f1}a. Following ultraviolet (UV) photoexcitation from the ground state (1A) to the electronic excited state (1B), the wave packet rapidly progresses along the steep 1B PES, inducing a stretching motion in the single Carbon-Carbon bond (C$_1$-C$_6$)~\cite{Kosma2009}, as illustrated in Fig.~\ref{f1}b. At the first CI (CI$_1$) between 1B and the dark intermediate state, the wave packet transitions to 2A~\cite{Kuthirummal2006}, recently reported as 3A~\cite{Travnikova2022}. Subsequently, the wave packet continues along the PES, approaching the second CI (CI$_2$) between the ground state and the intermediate state (pericyclic minimum)~\cite{Garavelli2001}. The wave packet eventually returns to the ground state and undergoes bifurcation at the pericyclic minimum, producing hot vibrational ring-closed CHD and ring-opened HT isomers (cZc, cZt and tZt) with varying torsions about the broken C-C bond~\cite{ruan_ultrafast_2001,Wolf2019}. The entire process, from photoexcitation to the formation of the initial ring-opened HT, occurs within approximately 150~fs~\cite{Sension2014,Pemberton2015,Minitti2015,Attar2017}. The recent few-fs ultrafast spectroscopic technique in the extreme UV regime has further revealed that the reaction occurs even faster (within 100 fs)~\cite{Karashima2021}. However, these studies are primarily sensitive to the electronic states, and the associated nuclear motions are not directly tracked. To date, the exact time required for the nuclear wave packet to traverse two CIs and their associated transient molecular structures remains elusive.

Ultrafast x-ray diffraction (UXD) and ultrafast electron diffraction (UED) complement spectroscopic techniques by directly probing molecular structures in momentum space rather than through transition energies, demonstrating their potential to directly track dynamic structures in real space and time~\cite{Centurion2022,Ischenko2017,Lee2024}. However, the requisite spatiotemporal resolution to resolve changes in molecular structure of CHD across the two CIs within 100 fs has not yet been achieved. Despite the superior temporal resolution offered by hard X-ray free-electron lasers, a direct real-space inversion was rarely used in UXD studies due to the limited momentum transfer range imposed by the limited wavelength~\cite{Minitti2015,Ruddock2019,Ma2020}. Consequently, conclusions were primarily drawn through comparisons with molecular dynamics simulations. In contrast, UED achieves a larger momentum transfer range due to the markedly short De Broglie wavelength of electrons. However, previous experiments resolved only the subsequent isomerization dynamics on the ground-state surface, while the initial excited-state ring-opening dynamics remained unresolved due to limited temporal resolution (>160~fs) of UED~\cite{ruan_ultrafast_2001,Ihee2001,Wolf2019}. Furthermore, both previous UXD and UED studies lacked information about the signature of the initially excited 1B state and its evolution upon photoexcitation.

In this work, by leveraging mega-electron-volt (MeV) UED to access a large momentum space (up to 10~$\text{\AA}^{-1}$) with enhanced temporal resolution (<80~fs), we provide key insights into the CI dynamics associated with the ring-opening reaction of CHD. In addition to the elastic scattering signal, which contains structural information, we observe inelastic scattering signals encoding information about the electronic excited state that are missing in previous measurements. This enables us to correlate nuclear dynamics with electronic dynamics, providing a comprehensive representation of the coupled non-adiabatic dynamics. Moreover, the diffraction-limited spatial resolution is surpassed by implementing a super-resolution inversion algorithm, allowing us to differentiate transient structures at the two CIs directly from experimental data. To validate the findings extracted from our analysis, we also compare the results with calculated ab initio non-adiabatic molecular dynamic trajectories as an independent comparison.

\section*{Results}

\begin{figure}
	\centering
	\includegraphics[width=1\textwidth]{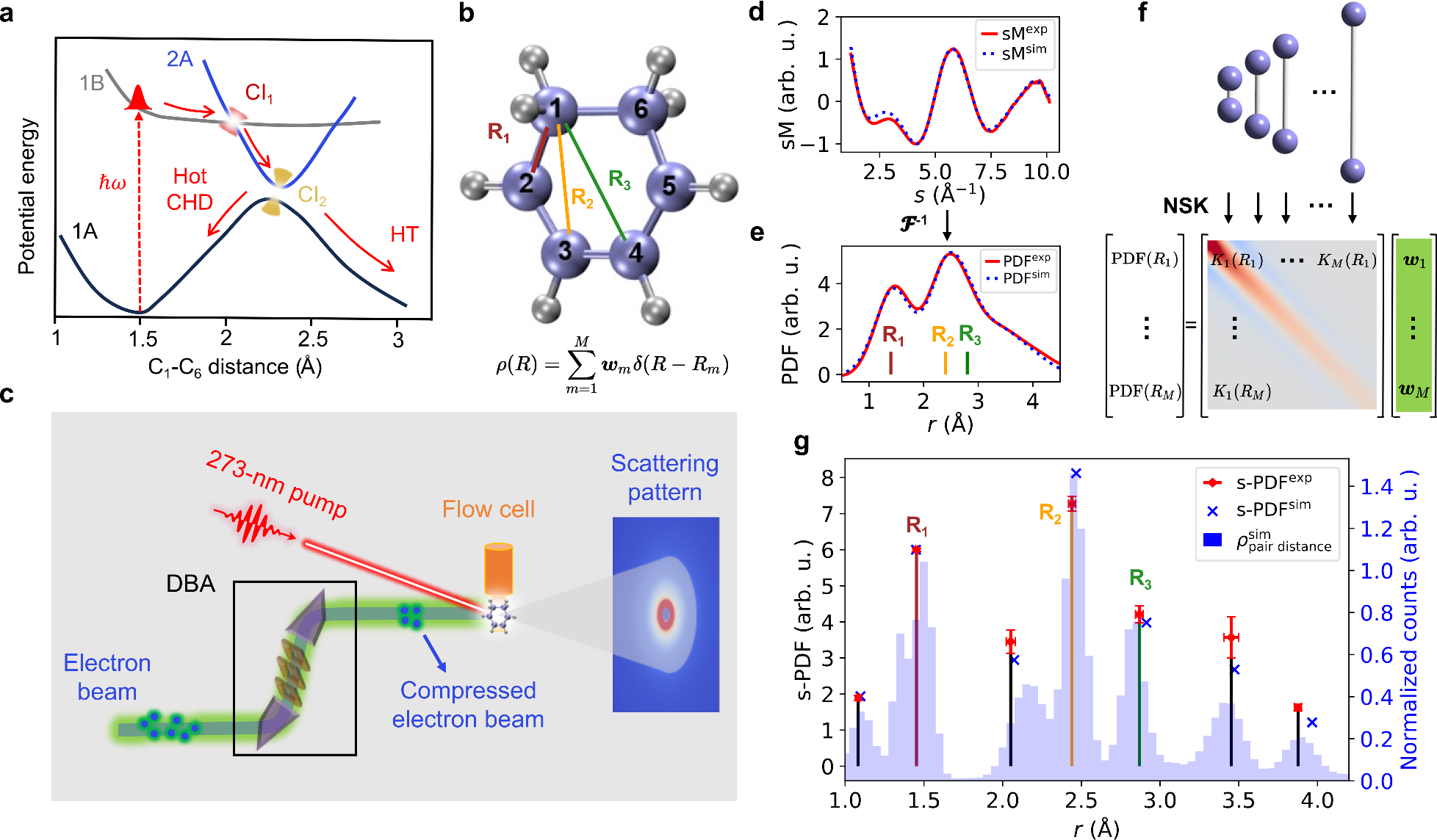} 
	\caption{Schematic of the reaction path, the experimental setup and the mechanism of super-resolution.
a, Reaction path and the relevant potential energy curves of ring-opening reaction of CHD. The red arrow indicates the direction of wave packet evolution. b, Static molecular structure. The closest, second nearest-neighbor, and diagonal C-C interatomic distances are designated as $R_1$, $R_2$, and $R_3$, respectively. The pair distance distribution in CHD can be expressed as a sum of weighted ($w_m$) $\delta$-functions.
c, Experimental setup. The focused 273-nm laser pulse intersects with the electron beam on the sample delivered via a flow cell. The electron beam is compressed by a double bend achromat (DBA) system to reduce both the electron pulse width and arrival time jitter. d, Experimental (red solid) and simulated (blue dotted) static modified molecular diffraction intensity (sM). e, Experimental (red solid) and simulated (blue dotted) static pair distribution function (PDF). The brown, orange and green lines represent $R_1$, $R_2$, and $R_3$, respectively.
f, Schematic of super-resolved real-space inversion. The natural scattering kernels (NSKs) are obtained by computing the PDF of each atomic pair with a given distance under the experimental constraints. The weights ($w$) to be reconstructed can be determined through convex optimization.
g, Experimental (red) and simulated (blue) s-PDFs. The light blue columns represent the interatomic distance distribution of the sampled static structures from the simulation, scaled with the scattering cross sections of the atomic pairs. The error bars correspond to one standard deviation calculated from a bootstrapped data set. Source data are provided as a Source Data file.
}
	\label{f1} 
\end{figure}

Our experimental setup is sketched in Fig.~\ref{f1}c. Briefly, a UV laser pump pulse (273~nm, 60~fs full width at half maximum, FWHM) is used to excite CHD molecules to the 1B state in gas phase, initiating the electrocyclic ring-opening reaction. An ultrashort electron beam (3 MeV) is then used to probe the reaction dynamics. The resulting diffraction patterns, produced by the scattering of electrons with the molecules, are recorded at different delay times between the UV pump pulse and the electron probe pulse. The ability to resolve both the two CIs and the evolution of the electronic excited state in our experiment is enabled by the enhanced temporal resolution of the MeV-UED instrument. This is achieved through the use of a double-bend achromat (DBA) lens, which reduces both the electron pulse width and timing jitter~\cite{Qi2020}. Unlike the conventional approach of mitigating Coulomb repulsion, which typically increases the electron pulse width, our instrument harnesses the Coulomb repulsion force to generate a positive linear chirp in the electron beam energy. Subsequently, the DBA lens with negative longitudinal dispersion is used to compress the elongated electron pulse. This method is similar to the well-established chirped pulse compression technique employed in laser metrology~\cite{Strickland1985}. As a result, the generated electron beam achieves a pulse width of approximately 40~fs (FWHM) and a timing jitter of around 20~fs (FWHM), yielding an instrument response function (IRF) of about 80 fs~(FWHM) after convolution with the pump laser (see Methods)~\cite{ma2022}.

As the scattering signal diminishes rapidly with increasing momentum transfer ($s$) of the scattered electron, the measured static scattering signal of ground-state CHD is represented by the modified molecular diffraction intensity sM$(s)$ (see Methods), as shown in Fig.~\ref{f1}d. The measurement is in excellent agreement with the simulated signal based on the sampled geometries according to the distribution function of the lowest vibrational level of the electronic ground state~\cite{Wigner1932,Crespo2018}. The effective $s$-range, $0.8<s<10\text{~\AA}^{-1}$, is approximately twice as large as that measured with UXD using 10 keV X-ray free-electron lasers~\cite{Minitti2015,Ruddock2019}. This extended range allows for the precise inversion of the data to yield a real-space distribution. The real-space pair distribution function (PDF), which represents the probability of finding an atomic pair at distance $r$, is obtained by performing a Fourier-sine transform on the static sM (see Methods). As illustrated in Fig.~\ref{f1}e, two broad peaks, centered at approximately 1.35~$\text{\AA}$ and 2.45~$\text{\AA}$, are clearly visible. The first peak corresponds to the nearest C-C interatomic distance $R_1$. The second peak is associated with both the second nearest-neighbor C-C interatomic distance $R_2$ and the diagonal C-C interatomic distance $R_3$, as indicated in Fig.~\ref{f1}b. Since the intensity of the PDF is proportional to the product of the scattering cross sections of the atom pairs, C-H and H-H interatomic distances do not produce distinct peaks in the PDF. Moreover, the diffraction limit results in a width of approximately $0.6~\text{\AA}$ for the PDF peaks. Consequently, $R_2$ cannot be distinguished from the diagonal C-C interatomic distance ($R_3\approx2.8~\text{\AA}$). Furthermore, the diffraction limit hinders the ability to discern small structural changes at the two CIs throughout the ring-opening reaction of CHD.

Inspired by super-resolution techniques in imaging and spectroscopy~\cite{Hell1994,Betzig2006,Moerner2015,Sigal2018}, we apply the recently proposed model-free deconvolution algorithm~\cite{Natan2023} to super-resolve atomic pairs with small differences in interatomic distance that are otherwise overlapped in conventional PDFs. 
Given that any atomic pair density distribution can be decomposed into a sum of weighted $\delta$-functions, the goal of super-resolution is to recover these weights from the measured PDF. As illustrated in Fig.~\ref{f1}f, the scattering kernels were obtained by computing the PDF of each $\delta$-function under the experimental constraints. The weights were then determined through convex optimization by minimizing the residual norm between the PDF constructed from the kernels and the measured PDF (see Methods and Supplementary Note 4).
The measured and simulated steady-state super-resolved PDF (s-PDF) of CHD are shown in Fig.~\ref{f1}g. In addition to clearly distinguishing the three types of C-C interatomic distances ($R_{1}$, $R_{2}$ and $R_{3}$), four types of C-H atomic pairs (see Supplementary Note 5), which are completely hidden in conventional PDF, are also resolved in s-PDF. Based on the simulated ground-state CHD, the statistical distributions of all interatomic distances are also obtained, as presented in Fig.~\ref{f1}g. The excellent agreement between the s-PDF and the statistical distribution of the pair distances provides strong evidence for the robustness and high accuracy of the super-resolution inversion algorithm. The combination of the super-resolution algorithm and the high temporal resolution of our UED instrument offers a new opportunity to capture the CI dynamics described in Fig.~\ref{f1}a.

\begin{figure} 
	\centering
	\includegraphics[width=1\textwidth]{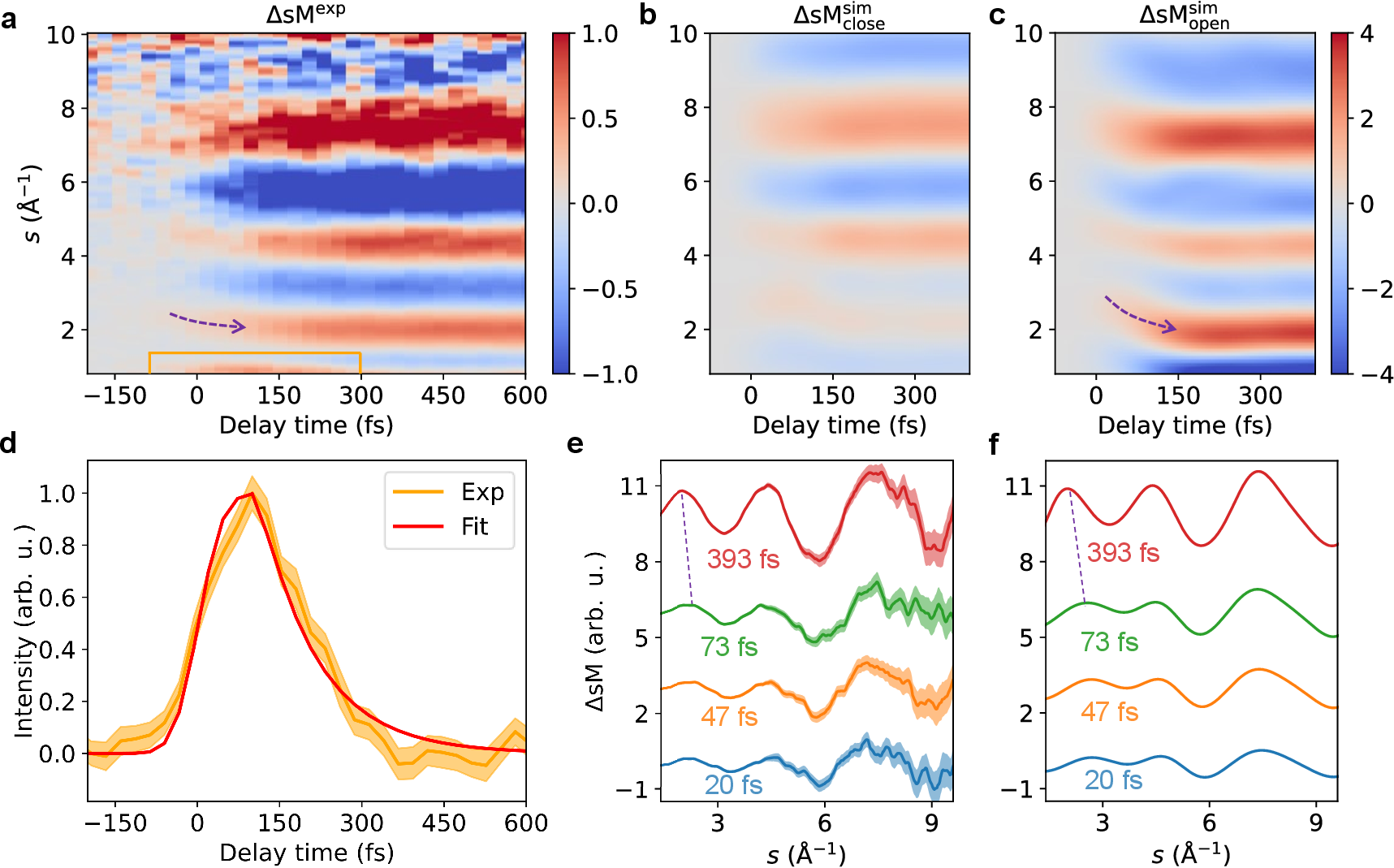} 
	\caption{Electronic and structural dynamics in momentum space.
    a, Experimental diffraction-difference signal in momentum space ($\Delta$sM) as a function of pump-probe delay. The orange box highlights the inelastic scattering signal and the purple dashed arrow indicates the position shift of the enhanced band at $\sim2\text{~\AA}^{-1}$. 
    b and c, The simulated $\Delta$sM for ring-closing and ring-opening trajectories, respectively. The simulated results are convolved with a Gaussian function with 80-fs FWHM. Two panels share a single colorbar. d, The integrated intensity of the inelastic signal. The fitting is described in the main text. e and f, The experimental and simulated $\Delta$sM at four selected time delays. The uncertainty represented by the shaded regions in d and e corresponds to a 68\% confidence interval. Source data are provided as a Source Data file.}
	\label{f2} 
\end{figure}

The measured time-resolved diffraction-difference signal in momentum space ($\Delta$sM, see Methods) at various time delays following photoexcitation is shown in Fig.~\ref{f2}a. As the measured diffraction patterns include contributions from both incoherent atomic scattering and coherent molecular interferences from interatomic pairs, and only a small fraction of molecules are excited, the diffraction-difference signal effectively eliminates the large contributions from unexcited molecules. This enables the detection of small signals related to structure changes that would otherwise be embedded in a large time-independent background. The measured diffraction-difference signal is characterized by positive and negative changes as a function of time delay and momentum transfer. To gain deeper insight into the characteristics exhibited by the experimental data, an independent ab initio surface-hopping molecular dynamic simulation was conducted at the XMS-CASPT2 level of theory \cite{Crespo2018} (see Methods). The simulation results are presented in Fig.~\ref{f2}b, c, showing the simulated $\Delta$sM for the ring-closing and ring-opening trajectories, respectively. For each channel, $\sim$100 trajectories were taken to provide the corresponding $\Delta$sM signal using the independent atom model (IAM)~\cite{Prince2006}. For ring-closing trajectories, the photoexcited molecules return to the ground state with a considerable amount of vibrational energy. These vibrationally hot molecules are characterized by significant fluctuations in interatomic distances, which lead to alterations in sM along $s$. Due to the inverse relation between momentum space and real space, the small distance changes in real space caused by the vibrationally hot CHD molecules result in $\Delta$sM predominantly exhibiting changes in the high momentum transfer region ($s>4~\text{\AA}^{-1}$), as illustrated in Fig.~\ref{f2}b. By contrast, as shown in Fig.~\ref{f2}c, considerable changes appear in the low momentum transfer region for the case of ring-opening trajectories, consistent with the large variation in interatomic distances as CHD molecules undergo conversion into ring-opened HT isomers. Furthermore, as indicated by the dashed arrow in Fig.~\ref{f2}c, the position of the enhancement band at approximately $s=2~\text{\AA}^{-1}$ shifts towards a lower value with increasing time delay during the first 150~fs, which is considered as a signature of ring-opening. Such a distinct feature is clearly observed in our experimentally measured $\Delta$sM (indicated by the dashed arrow in Fig.~\ref{f2}a).

Besides the three enhanced bands and four bleached bands observed in Fig.~\ref{f2}a, which are in excellent agreement with the simulation, a strong enhancement signal was observed in the very low momentum transfer region ($s<1.2~\text{\AA}^{-1}$, as indicated by the orange rectangle) immediately after photoexcitation. This signal, however, is absent in the simulation based on the IAM model. The anomalous signal persists for only a brief period, in stark contrast to the signal observed in the region $1.2<s<10~\text{\AA}^{-1}$, which is associated with the structural changes in the molecule and persists throughout the entire measurement duration. This low momentum transfer signal is due to small-angle inelastic scattering, which is linked to changes in the electronic states that have been previously observed in other systems~\cite{Yang2020,Champenois2023,wang2025,green2025}. The normalized integrated small-angle scattering signal is shown in Fig.~\ref{f2}d (see Supplementary Note 6). The sharp rising edge of the signal is fitted with an error function with a width corresponding to the experimental IRF. Given the difficulty of performing an in-situ cross-correlation between the UV laser and the electron beam, the time zero of the measurement is defined as the center of the IRF determined in the aforementioned fitting. The falling edge of the signal is not resolution-limited and can be modeled with a convolution of the IRF and an exponential decay, yielding a decay constant of $100\pm7$~fs. This is attributed to the lifetime of the initially excited electronic state and is comparable to values reported in previous spectroscopic measurements~\cite{Attar2017}. This supports the hypothesis that the small-angle scattering signal originates from changes in the electronic excited state. It is worth noting that the inelastic scattering signal for CHD has not been observed in previous UED experiments, likely due to limited temporal resolution and signal-to-noise ratio~\cite{ruan_ultrafast_2001,Ihee2001,Wolf2019}. In the context of ultrafast X-ray scattering with X-ray free-electron lasers, while the temporal resolution is not a limiting factor, the cross section of inelastic scattering is much lower at small momentum transfer~\cite{Ma2020}. Consequently, this signal was not observed in previous ultrafast X-ray scattering experiments~\cite{Minitti2015,Ruddock2019}. 

As the measured $\Delta$sM is a combination of contributions from both the ring-closing and ring-opening reactions, it can be used to estimate the branching ratio of the two relaxation pathways. The best fit yielded a quantum yield of $32\pm3\%$ for the HT isomers (see Methods), which is comparable to the predicted value from the simulation (about $42\%$) and to the measured value obtained through spectroscopic methods (about $30\%$)~\cite{Adachi2015}. It is noteworthy that the branching ratio derived from the fitting is lower than the value (about $50\%$) reported in a previous UED measurement~\cite{Wolf2019}. This discrepancy could be ascribed to the slightly different central wavelengths of the pump pulses (267~nm vs 273~nm) used in the two experiments. To gain further insight into the structural dynamics, the measured values of $\Delta$sM at four representative time delays are presented in Fig.~\ref{f2}e. The simulated $\Delta$sM with the fitted branching ratio, as shown in Fig.~\ref{f2}f, is in excellent agreement with the measurements. Specifically, at the earliest stage, prior to the opening of the ring, the most pronounced signal is the depletion observed at $s\approx6~\text{\AA}^{-1}$, as the signal at large momentum transfer is generally sensitive to structural changes at short distances. In contrast, the signals at $\sim$2~$\text{\AA}^{-1}$ and $\sim$3~$\text{\AA}^{-1}$ only gain a significant increase after the wave packet has traversed the second CI, as the signal at small momentum transfer is generally sensitive to structural changes at long distances. Furthermore, the position of the peak at $\sim$2~$\text{\AA}^{-1}$ exhibits a notable shift towards the low $s$ region after 100 fs, indicating the onset of the ring-opening after the second CI, which eventually leads to the formation of HT.

\begin{figure} 
	\centering
	\includegraphics[width=0.8\textwidth]{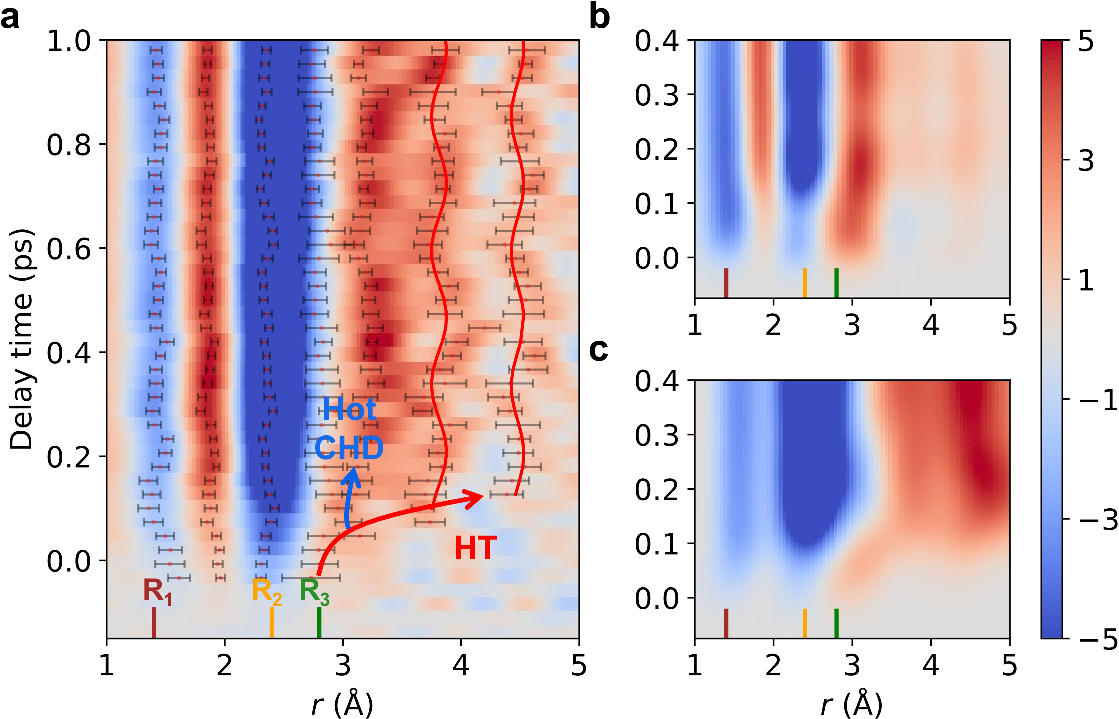} 
	\caption{Photoinduced ultrafast dynamics visualized in real space.
a, The background density map corresponds to the conventional $\Delta$PDF derived from measured $\Delta$sM, while the positions of the sparse pair distributions reconstructed through $\Delta$s-PDF are marked as scatter points overlaid on the $\Delta$PDF. The error bars are estimated via bootstrap resampling and represent one standard deviation. The closest ($R_1$), second nearest-neighbor ($R_2$), and diagonal ($R_3$) C-C interatomic distances for the steady-state structure are given at the bottom. The wave packet bifurcates at approximately 100~fs and the subsequent evolution in the ring-opening and ring-closing pathways are indicated by the arrows. The center of the enhanced bands for HT is determined with super-resolution analysis. b and c, Simulated $\Delta$PDF of ring-closing and ring-opening trajectories, respectively. The simulated results are convolved with a Gaussian function with 80-fs FWHM. Source data are provided as a Source Data file.
}
	\label{f3} 
\end{figure}

A distinctive advantage of UED is its capability to access a large momentum transfer range, enabling precise inversion to structural evolution in real space. Figure~\ref{f3}a-c show the measured and simulated difference PDFs ($\Delta$PDFs) obtained from a Fourier-sine transform of $\Delta$sM in Fig.~\ref{f2}a-c after removing the inelastic scattering contribution at small scattering angles (see Supplementary Note 3), while the positions of the sparse pair distribution reconstructed from the $\Delta$s-PDF are marked as scatter points overlaid on the $\Delta$PDF in Fig.~\ref{f3}a (see Methods). As shown in Fig.~\ref{f3}b, the vibrationally hot CHD in the ring-closing pathway exhibits significant fluctuation in the interatomic distances. This results in the bleaching of bands at $r\approx R_1$ and $r\approx R_2$ and the enhancement of bands at slightly larger interatomic distances, e.g. $r\approx1.9~\text{\AA}$ and $r\approx3~\text{\AA}$. Regarding the ring-opening pathway, it is notable that the enhanced bands observed at $r\approx1.9~\text{\AA}$ and $r\approx3~\text{\AA}$ are absent in Fig.~\ref{f3}c, despite the presence of bleached bands at similar distances caused by the lengthening of the $R_1$, $R_2$ and $R_3$ distances in CHD. Additionally, a considerable intensification of the signal was observed within the range $3.5 <r<5~\text{\AA}$, which provides definitive evidence for photoinduced ring-opening. All these features are observed in the measurement, as illustrated in Fig.~\ref{f3}a. In particular, three enhanced bands appear in the region $3<r<5~\text{\AA}$, with the first band around 3 $\text{\AA}$ from the hot CHD and the other two bands at 3.8~$\text{\AA}$ and 4.5~$\text{\AA}$ corresponding to the HT isomers (see Methods and Supplementary Fig.~4). The centers of the bands for HT determined by $\Delta$s-PDF exhibit discernible oscillations with a period of approximately 300~fs, as indicated by the red solid lines. These oscillations can be attributed to interconversion between HT isomers due to the rotation of the terminal ethylene groups~\cite{Wolf2019}. A detailed examination of the signals at the initial stage reveals the presence of a bifurcation point for the wave packet at approximately 100~fs, beyond which the two enhanced bands at $r>3.5~\text{\AA}$ rapidly emerge. This suggests that the wave packet reaches the pericyclic minimum approximately 100~fs following photoexcitation. This finding is further corroborated by the inelastic signal, which reached its maximum value at $t\approx100$~fs as shown in Fig.~\ref{f2}d.

Although the bifurcation at CI$_2$ can be observed through the analysis of the $\Delta$PDFs (Fig.~\ref{f3}a), which mirrors the findings observed in the case of CF$_3$I~\cite{Yang2018}, such protocol cannot be applied to locate the first CI due to the absence of a pronounced bifurcation of the structural pathway in the observed $\Delta$PDFs. Consequently, it is imperative to develop new signatures for the extraction of CI dynamics, as opposed to merely observing bifurcations. It is important to note that the observed bleached band at $r\approx2.5~\text{\AA}$ is a consequence of changes in the distances between both the second-nearest C-C ($R_2$) and the diagonal C-C ($R_3$) atomic pairs. However, the pair density changes around two interatomic distances could not be extracted directly using the conventional PDF analysis. This limitation has thus far proven to be an insurmountable obstacle in capturing CI dynamics in real space, where the changes in distance are smaller than those permitted by the diffraction limit.
%However, the $\Delta$s-PDF successfully resolves the peak for $R_3$ at $\sim$2.8~$\text{\AA}$, which is indistinguishable from $R_2$ at $\sim$2.4~$\text{\AA}$ in Fig.~\ref{f3}a.
Notably, the $\Delta$s-PDF suggests a peak near $R_3$ at $\sim$2.8~$\text{\AA}$, distinct from $R_2$ at $\sim$2.4~$\text{\AA}$ in Fig.~\ref{f3}a, as a result of the sparsity-based reconstruction. We note that in the case of broadened or delocalized nuclear wave packets, the resulting $\Delta$s-PDF could exhibit multiple adjacent discrete points representing a broad continuous distribution, which does not necessarily indicate a well-separated atom pair at that specific distance. Nevertheless, even when the sparsity condition is compromised, the $\Delta$s-PDF still serves as a practical tool for robustly deconvolving the conventional $\Delta$PDF and extracting localized variations in pair density.
In the early stages of the reaction, atomic pairs shorter than R$_3$ extend through the R$_3$ distance during the ring-opening process. This elongation masks the depletion of the diagonal C-C interatomic distance, resulting in a positive peak around R$_3$. Consequently, the negative peak around R$_3$, which arises from the depletion, becomes apparent after the system has undergone ring opening (beyond $\sim$100~fs).
%Thus, the $\Delta$s-PDF is capable of resolving the small difference in interatomic distances beyond the diffraction limit. 
Here, we further demonstrate that the combination of $\Delta$s-PDF and the high temporal resolution of the experimental setup can capture the subtle structural changes that allow the two CIs to be located with direct real-space inversion imaging. Our analysis reveals that the $\Delta$s-PDF exhibits a positive peak at approximately $1.95~\text{\AA}$ and a negative peak at approximately $2.30~\text{\AA}$ immediately following photoexcitation. Given the calculated C$_1$-C$_6$ distances of approximately $2~\text{\AA}$ at CI$_1$ and $2.2~\text{\AA}$ at CI$_2$, we attribute the observed pair density changes extracted through $\Delta$s-PDF at $1.95~\text{\AA}$ and $2.30~\text{\AA}$ to the nuclear wave packet motion near two CIs.

\begin{figure}
	\centering
	\includegraphics[width=0.8\textwidth]{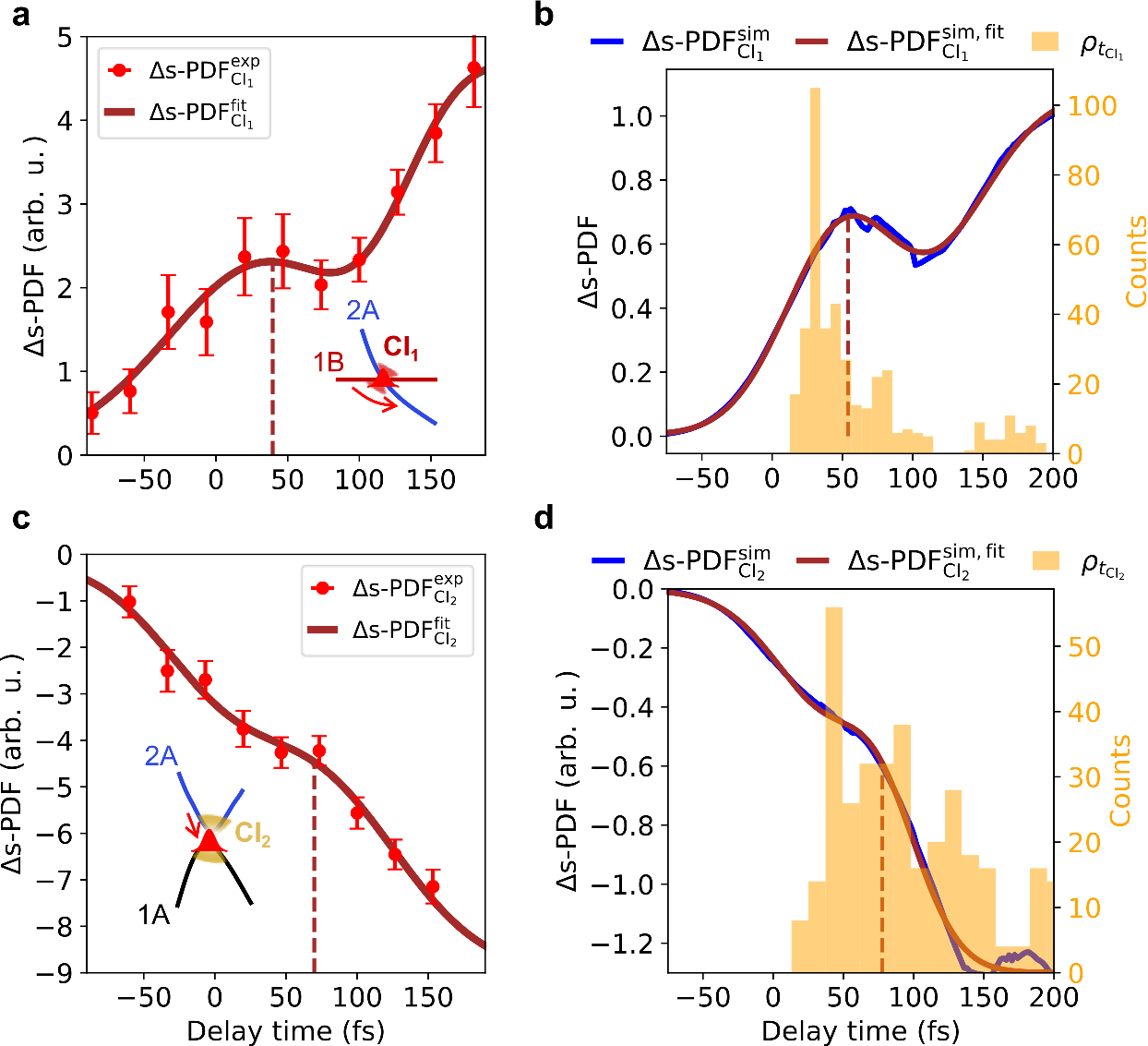} 
	\caption{Structural dynamics during passage through the two CIs.
    a and c, Measured intensity of $\Delta$$\text{s-PDF}$ at $\sim$1.95$\text{~\AA}$ and $\sim$2.30$\text{~\AA}$, respectively. The brown solid lines show the results of least-squares fitting of the data with a sum of Gaussian function and error function. The brown dashed line indicates the center of the Gaussian function obtained from the fitting. The error bars correspond to one SD calculated through a bootstrapped data set. b and d, Simulated $\Delta$$\text{s-PDF}$ at $\sim$1.95$\text{~\AA}$ and $\sim$2.30$\text{~\AA}$, respectively. The orange bar graphs are the distributions of $t$$_{\text{CI}_1}$ and $t$$_{\text{CI}_2}$ derived from 193 trajectories in surface hopping simulations. The simulated results are convolved with a Gaussian function with 80-fs FWHM. Source data are provided as a Source Data file.
    }
	\label{f4} 
\end{figure}

Figure~\ref{f4}a illustrates the measured evolution of the $\Delta$s-PDF at $1.95~\text{\AA}$. In the absence of an atomic pair at this distance prior to photoexcitation (see Fig.~\ref{f1}g), the initial increase in intensity suggests that the wave packet motion is associated with the lengthening of the C$_1$-C$_6$ interatomic distance. After this initial rise, the intensity undergoes a slight decline at $t\approx50$~fs, followed by a subsequent increase at $t\approx100$~fs. The observed decline can be attributed to the wave packet sliding away as it passes through the first CI, while the subsequent rise corresponds to the formation of the ring-closed CHD, where the C$_1$-C$_6$ distance decreases. A fit to the data indicates that the wave packet reaches the first CI at approximately $40\pm17$~fs after photoexcitation. The simulated $\Delta$s-PDF exhibits a similar evolution, with a local maximum occurring at approximately 54~fs. Furthermore, the statistical distribution of the wave packet arrival time at the first CI  ($t_{\text{CI}_{1}}$), which was obtained from the simulated trajectories by minimizing the energy difference between S$_1$ and S$_2$, is shown in Fig.~\ref{f4}b. The mean value of $t_{\text{CI}_{1}}$ is found to be approximately 50~fs, closely aligning with the experimental result derived from super-resolved structural dynamics.

Figure~\ref{f4}c illustrates the measured evolution of the $\Delta$s-PDF at $2.30~\text{\AA}$. Given that the atomic pair density has a notable value at this distance (see Fig.~\ref{f1}g) prior to photoexcitation, primarily due to the contribution of the second nearest C-C interatomic distance $R_2\approx2.4~\text{\AA}$, the lengthening of the C$_1$-C$_6$ distance following photoexcitation results in a further increase in $R_2$ and a reduction in its contribution to the s-PDF at $2.30~\text{\AA}$. Consequently, the measured intensity exhibited a pronounced decline upon photoexcitation. The intensity decrease then slowed down, and a knee structure was observed, followed by a subsequent decline. This knee structure can be attributed to the interplay between the nuclear wave packet motion associated with C$_1$-C$_6$ and the second nearest C-C distances. Specifically, the C$_1$-C$_6$ distance increases to approximately $2.2~\text{\AA}$ as the wave packet approaches the second CI. This resulted in an enhancement of the intensity of the $\Delta$s-PDF at $2.30~\text{\AA}$, partially compensating for the reduction caused by the lengthening of $R_2$ and slowing the overall decline. After passing through the second CI, the C$_1$-C$_6$ distances for both the ring-opening and ring-closing pathways diverge from this region, initiating the subsequent decline after the knee structure. A fit to the data indicates that the wave packet arrives at the second CI at approximately $69\pm18$~fs following photoexcitation. The simulated $\Delta$s-PDF also exhibits a comparable knee structure at 78~fs, which aligns closely with the mean hopping time from $S_1$ to $S_0$ of the wave packet at the second CI ($t_{\text{CI}_{2}}$), as shown in Fig.~\ref{f4}d. Moreover, the extracted CI$_2$ passage time closely matches the onset of the decay in the inelastic scattering signal, indicating simultaneous changes in both the nuclear and electronic degrees of freedom. This temporal coincidence provides further evidence that the observed dynamics are governed by the conical intersection. The results of our s-PDF analysis, which provide information on the density distributions associated with wave packet motion, reveal that the nuclear wave packet traverses from CI$_1$ to CI$_2$ in only approximately 30~fs. This is in close agreement with the time obtained from the theoretical simulations. It should be noted that it is not uncommon to resolve a time-delay smaller than the IRF of the measurement~\cite{Attar2017,Yang2018} (see Supplementary Note 8). 

Although the traversal time of the nuclear wave packet through CIs can be predicted from theoretical simulations, determining it directly from experiments is highly challenging. However, the unique capability of super-resolved UED makes this possible, as it can identify distinct transient structures at two different CIs along with the timing of their occurrence. This can provide new benchmarks for quantum chemistry methods and molecular dynamics simulations. It is interesting to note that if a constant speed is assumed, the separation speed of the C$_1$-C$_6$ bond is estimated to be approximately 12 Å/ps. Such rapid and forceful motion can quickly push aside neighboring molecules, which is why these ultrafast electrocyclic reactions are ideal candidates for artificial molecular machines~\cite{Eelkema2006,Erbas2015}.

In summary, the analysis of elastic and inelastic scattering signals using ultrashort electron pulses, supported by ab initio molecular dynamics simulation, provides valuable insights into the key nuclear and electronic dynamics involved in the ring-opening reaction of CHD. Furthermore, it has been demonstrated that diffraction-limited spatial resolution can be circumvented by applying a super-resolution inversion algorithm. Using super-resolution femtosecond electron diffraction, we show that the subtle structural changes in CHD molecules during the early stages of photoinduced ring-opening dynamics can be resolved. By leveraging the ability to measure density distributions associated with nuclear wave packet motion, we reveal that it takes approximately 30~fs for the nuclear wave packet to traverse the two CIs. The demonstrated super-resolution technique should be applicable to both UXD and UED to enhance the spatial resolution. With ongoing improvements in temporal resolution and signal-to-noise ratio, super-resolution ultrafast scattering techniques hold great promise for advancing photochemistry and photophysics by spatially resolving intricate coupled electronic and nuclear motions originating from transient electronic coherence at CIs~\cite{Rouxel2021,Rouxel2022} with unparalleled precision. This could fill a critical gap in ultrafast molecular science, potentially unlocking pathways for faster electronics, advanced materials, and quantum computing through the manipulation of molecular properties at the most fundamental level.

%%%%%%%%%%%%%%%% MAIN TEXT FIGURES %%%%%%%%%%%%%%%
%\makeatletter
%\renewcommand{\fnum@figure}{\textbf{Fig. \thefigure}}
%\renewcommand{\fnum@table}{\textbf{Table \thetable}}
%\makeatother

%%%%%%%%%%%%%%%% METHODS %%%%%%%%%%%%%%%

%\makeatletter
%\renewcommand{\fnum@figure}{\textbf{Extended Data Fig. \thefigure}}
%\renewcommand{\fnum@table}{\textbf{Extended Data Table \thetable}}
%\makeatother
%\setcounter{figure}{0}
%\setcounter{table}{0}

\clearpage

\section*{Methods}

\subsection*{Gas-phase MeV UED }
The gas-phase MeV-UED instrument is schematically shown in Supplementary Fig.~1. The instrument utilizes a Ti:sapphire laser system (Vitara and Legend Elite Duo HE, Coherent) with a center wavelength of approximately 820~nm and a pulse width of about 30~fs (FWHM) to generate both the UV pump laser pulse and the electron probe pulse. The electron beam is generated in a 2.33-cell photocathode RF gun (2856 MHz), where the intense RF field rapidly accelerates the electron beam to relativistic velocity ($v\approx0.99$~c) with a kinetic energy of about 3~MeV. During the acceleration and propagation stages, the electron beam's pulse width is increased to a few hundred femtoseconds due to the Coulomb repulsion force. This also produces a positive energy chirp in the beam's longitudinal phase space. A double-bend achromat (DBA) lens with a negative longitudinal dispersion, consisting of two dipole magnets and three quadrupole magnets, is further used to compress the electron pulse width to approximately 30~fs (FWHM). The beamline is meticulously optimized to ensure the isochronous transport of the electron beam from the cathode to the sample. This approach significantly reduces the sensitivity of the electron beam's time of flight to fluctuations in both the amplitude and phase of the radio-frequency field, resulting in a minimal timing jitter of approximately 20~fs (FWHM). 

The CHD sample is purchased from Adamas without further purification. The gas is delivered with a flow cell measuring 2~mm in length and 0.7 mm in diameter. The gas cell is heated to $\sim$85~$^\circ$C to prevent sample condensation. The electron beam size at the gas cell is measured to be $\sim$150~µm (FWHM). The UV pump laser has a center wavelength of 272.6~nm with a bandwidth of 1.8~nm (FWHM). The laser spot size at the gas cell is approximately 300~$\times$~320~µm (FWHM) and the pulse energy is set within the linear absorption range of CHD (see Supplementary Note 2). The UV laser intersects the electron beam at the sample with a small angle ($\sim$$3^\circ$), ensuring a small velocity mismatch between the pump and the probe pulses. The temporal resolution can be calculated as
$\tau_{{\rm{total}}}=\sqrt{\tau_{{\rm{pump}}}^2+\tau_{{\rm{probe}}}^2+\tau_{{\rm{jitter}}}^2+\tau_{{\rm{VM}}}^2}$, where $\tau_{{\rm{pump}}}$ is the pulse width of the pump laser, $\tau_{{\rm{probe}}}$ is the pulse width of the electron beam, $\tau_{{\rm{jitter}}}$ is the timing jitter and $\tau_{{\rm{VM}}}$ is the degradation of the resolution caused by the velocity mismatch. With the laser pulse width being approximately 60~fs (FWHM) and $\tau_{{\rm{VM}}}\approx40~$fs, the temporal resolution, which is taken as the instrument response function (IRF) of the experiment, is estimated to be approximately 80~fs (FWHM).  

The diffraction pattern is measured with a phosphor screen imaged to an electron-multiplying charge-coupled device (EMCCD). The phosphor screen is designed with a central aperture measuring 3 millimeters in diameter, permitting the transmission of unscattered electrons. Consequently, the data within the range of $s<0.8~\text{\AA}^{-1}$ are not measured. The repetition rate of the electron beam is 200 Hz and each scattering pattern at a designated time delay is accumulated over a period of 14 seconds with 2800 electron pulses, each with an approximate charge of 20~fC. The diffraction pattern is measured with a time step of 26.7~fs in the time window from -0.4~ps to 1.2~ps. The full data sets include 308 such scans. 

\subsection*{Data processing}
Our data processing procedures are similar to previous works~\cite{Ihee2002,Yang2020}. Briefly, all 2D images at the same time step are normalized and averaged, then the isotropic 1D scattering signal $I(s,t)$ is extracted. The difference scattering intensity is computed by ${\Delta{I}}(s,t)=I(s,t)-I(s,t<0)$. Both the background and the atomic signal are subtracted without further processing. Supplementary Figure~2 shows the measured percentage difference (PD) signal. Then the modified difference scattering intensity is calculated by $\Delta {\rm{sM}}(s,t)=\frac{{\Delta{I}}(s,t)}{I_a}s$ (Supplementary Note 1).

\subsection*{Conventional real-space inversion}
The conventional PDF ($\Delta$PDF) can be computed by performing a Fourier transform on the measured sM ($\Delta$sM):
\begin{equation}
    {\rm{PDF}}(r)=r\int_{s_{min}}^{s_{max}} {\rm{sM}}(s)\sin(sr)e^{-ks^2}ds,
    \label{cal_PDF}
\end{equation}
\begin{equation}
    \Delta {\rm{PDF}}(r)=r\int_{s_{min}}^{s_{max}} \Delta {\rm{sM}}(s)\sin(sr)e^{-ks^2}ds.
    \label{delta_PDF}
\end{equation}
The ${\rm{PDF}}(r)$ is employed to compute the static molecular structure, while the $\Delta {\rm{PDF}}(r)$ is utilized to analyze time-resolved structural change. The function $e^{-ks^2}$  is used to prevent non-physical signals caused by truncation in the momentum space and to mitigate the effect of increased noise at high-$s$ region. In this study, $k$ is set to 0.03~$\text{\AA}^2$ and $s_{max}$ is taken to be $10~\text{\AA}^{-1}$. It should be noted that we need to remove the inelastic scattering contribution to retain only the elastic scattering signal when performing the conventional PDF transformation (more details in Supplementary Note 3).

\subsection*{Super-resolved real-space inversion}
The s-PDF is obtained using a model-free inversion technique~\cite{Natan2023} that has been recently developed. Given the distribution of any arbitrary atomic pairs can be decomposed into a sum of $\delta$-functions:
\begin{equation}
    \rho(R)=\sum_{m=1}^{M}w_m\delta(R-R_m),
    \label{rho_weights}
\end{equation}
we can decompose the measured PDF into a sum of weighted PDFs calculated from different $\delta$-functions.
Thus, we compute the PDF generated by each $\delta(R-R_m)$ distribution under the experimental conditions (scattering kernels) and we combine these scattering kernels into a dictionary ($\mathcal{D}$). Then the weights in Eq.~(\ref{rho_weights}) can be determined through convex optimization by minimizing the residual norm between the dictionary constructed from the kernels and the measured PDF:
\begin{equation}
    ||\Delta {\rm{PDF}}-\mathcal{D}\boldsymbol{w}||^2 +\alpha \mathcal{R}(\boldsymbol{w}).
\end{equation}
Here, $\mathcal{R}(\boldsymbol{w})=\sum_{m}|\boldsymbol{w}_m|$ is the $l_1$ regularizer~\cite{Tibshirani1996} in convex optimization. The validity of the deconvolution can be further substantiated by employing the L-curve method~\cite{Hansen1993}. More details on super-resolved real-space inversion are provided in Supplementary Note 4.

The model-free super-resolution reconstruction algorithm is fundamentally based on the theory of compressed sensing~\cite{Candes2006}, where the reconstruction accuracy of the super-resolution approach is influenced by the degree of sparsity. In the event that the distribution is characterized by concise representations when expressed in a proper basis, it is possible to accurately recover the sparse distribution beyond the diffraction limit. In comparison to alternative methods that depend on global optimization and exhaustive structure refinement, the super-resolution method exhibits superior efficiency and reduced computational demands. However, for atomic pair distributions that are not sufficiently sparse, the reconstruction accuracy is known to decrease. This issue can be resolved through a transformation of the distribution into a different basis, thereby rendering it sparse. However, it should be noted that such transformations frequently depend to a significant extent on prior knowledge. Additionally, while super-resolution can facilitate the recovery of atomic pair distributions, it does not inherently yield detailed molecular structural information, such as angular configurations in three-dimensional space. In this work, the critical real-space information is found in the evolution of the C$_1$–C$_6$ atomic pair distance, which can be reconstructed using the super-resolution method.

\subsection*{Geometries simulations}
The geometry optimization and frequency analysis of the ground state minima (S$_0$-min) of CHD are performed at the MP2 level with the def2-TZVP basis using the Gaussian 16 package~\cite{Gaussian2016}. Three photoproducts (HT isomers), cZc-HT, cZt-HT and tZt-HT, are also optimized at the MP2/def2-TZVP level.

The single point calculation at S$_0$-min was performed with the multistate complete-active-space second-order perturbation theory (XMS-CASPT2)~\cite{Shiozaki2011}. The active space contains 6 electrons in 6 orbitals (6e, 6o), which contains the two $\pi$ orbitals and their antibonding $\pi^*$ orbitals, as well as the $\sigma$ and $\sigma^*$ orbitals that located at the C$_1$-C$_6$ bond, see Supplementary Fig.~3. To avoid possible intruder states in the XMS-CASPT2 calculations, the real level shift of $0.5E_h$ is employed~\cite{Roos1995}. The absence of convergence failures due to intruder states in the subsequent nonadiabatic simulations further confirms the robustness of the chosen level shift value. In all XMS-CASPT2 calculations, the orbital and density fitting basis sets are chosen as def2-TZVP and def2-universal-JKFIT, respectively~\cite{Weigend2005}. All XMS-CASPT2 calculations are performed with the three-state average using the BAGEL program~\cite{Park2017}. 

In addition, two conical intersection geometries of S$_2$-S$_1$ (CI$_1$) and S$_1$-S$_0$ (CI$_2$) are optimized at the same XMS-CASPT2 level, see Supplementary Fig.~4b, c. 
The optimized geometries of the S$_0$-min, CI$_1$ and CI$_2$ are consistent with previous works~\cite{Wolf2019}. The energies and electronic characters of low-lying electronic states at these geometries are given in Supplementary Table~1 and Supplementary Table~2. The optimized ring-opened geometries of three HT isomers (cZc, cZt and tZt) are shown in Supplementary Fig.~4d-f. The stretched C-C interatomic distances contribute signals at $r>3.5~\text{\AA}$ in PDF during the ring-opening dynamics.

\subsection*{Preparation of initial conditions}
5000 initial conditions (nuclear coordinates and velocities) are sampled from the Wigner distribution function of the lowest vibrational state at the S$_0$-min~\cite{Crespo2018}. The electronic excited-state characters including vertical excitation energies and oscillator strengths of all sampled geometries are obtained at the XMS (3)-CASPT2 (6, 6)/def2-TZVP level. Then these 5000 snapshots are taken to calculate the absorption spectra with the nuclear ensemble approach (NEA)~\cite{Crespo2012}, and the stick spectra are broadened by the Lorenz line shape, in which the broadening parameter is chosen as 0.05~eV. Given the pump laser pulse, a narrow excitation energy window with $\sim$2~nm FWHM (4.39-4.46~eV) is employed to choose the initial geometries in the trajectory surface hopping dynamics. The center of the energy window in the simulation is shifted by comparing the energy difference between the peak position of the experimental~\cite{Kosma2009} and simulated absorption spectra~\cite{Hait2024}. 193 initial conditions within this energy window are selected to start the nonadiabatic dynamics from S$_1$.

\subsection*{Trajectory surface hopping simulations}
The ultrafast nonadiabatic dynamics simulations of CHD are calculated with the on-the-fly trajectory surface hopping (TSH) method proposed by Tully’s fewest-switches algorithm at the XMS(3)-CASPT2(6, 6)/def2-TZVP level~\cite{Crespo2018}. The simulation time window is 0-400~fs. The nuclear motions are propagated with a time step of 0.5~fs, with 100 electronic propagation substeps for each nuclear step. The decoherence correction approach proposed by Granucci et al.~\cite{Granucci2007} is employed, and the parameter is set to 0.1 Hartree~\cite{Zhu2004}. At hops, the nuclear velocities are corrected along the direction of the nonadiabatic coupling vector to satisfy the energy conservation. For frustrated hops, the components of the velocities are reversed along the direction of the nonadiabatic coupling vector.
All nonadiabatic dynamics simulations are performed with the JADE-NAMD package~\cite{Hu2021} that combines the BAGEL program, which contains the interface between nonadiabatic dynamics and electronic calculations.

A total of 193 trajectories are used to analyze the structural dynamics. The diffraction signals of the simulated nuclear structures are computed with IAM model by
\begin{equation}
    I_{m}(s)=\sum_{i=1}^{N}\sum_{j=1,j\neq i}^{N}f_i^*(s)f_j(s)\frac{\sin(sr_{ij})}{sr_{ij}}.
\end{equation}
Here, $f_i(s)$ is the elastic scattering amplitude for the $i$th atom and is calculated by the ELSEPA program~\cite{Salvat2005}.

\subsection*{Quantum yield and excitation ratio estimation}
In view of the substantial difference in $\Delta$sM for ring-closing and ring-opening trajectories,  we leverage this distinctive feature to estimate the quantum yield of HT in our experiment. This is achieved by comparing the measured $\Delta$sM with the simulated results obtained at varying quantum yields. Simulations corresponding to different quantum yields are obtained by resampling from the set of 193 computed trajectories, in which a specified number of ring-opening and ring-closing trajectories are randomly selected to match the desired yield. To ensure statistical convergence, multiple resampling iterations are performed, and as many trajectories as possible are included in the averaging process. The $\Delta$sM are averaged over 300-400~fs, at which point the ring opening dynamics are complete, for both the experimental $\Delta$sM$^{\text{exp}}_{\text{ave}}$ and theoretical results $\Delta$sM$^{\text{sim}}_{\text{ave}}$. The residual norms of $\Delta$sM$^{\text{exp}}_{\text{ave}}$-$\Delta$sM$^{\text{sim}}_{\text{ave}}$ at $1.5<s<8~\text{\AA}^{-1}$ with different quantum yields for ring-opened HT are shown in Supplementary Fig.~5. The best fit yields a quantum yield of $32\pm3~\%$ for the HT. The simulated results with the adjusted branching ratio can thus be regarded as a kinetic fit based on selected trajectories, providing a useful framework for analyzing the ring-opening dynamics.

Based on the obtained quantum yield, the excitation ratio can be estimated by
\begin{equation}
    p_{\rm{exc}}=\frac{\rm{PD^{exp}}}{\rm{PD^{sim}_{100\%}}},
\end{equation}
where $\rm{PD^{exp}}$ is the measured PD, $\rm{PD^{sim}_{100\%}}$ is the simulated PD with 100\% excitation ratio with $32~\%$ ring-opening ratio. The result suggests that approximately 8\% of the scattered molecules are excited.

\subsection*{Simulated population and nuclear dynamics}
As illustrated in Supplementary Fig.~6, upon initiating the nonadiabatic dynamics from the S$_1$ state, the system rapidly decays to S$_0$ with a time constant of 84~fs. Moreover, it is observed that more than half of the trajectories decay to S$_0$ within 100~fs, with a confidence level of 96.4\%. These results are consistent with previous results~\cite{Wolf2019}. However, it is important to note that the changes in electronic state populations do not directly give detailed information about the structural dynamics.

Alternatively, the ring-opening and ring-closing structural dynamics can be readily visualized through the evolution of the C$_1$-C$_6$ distance. The evolution of the C$_1$-C$_6$ bond distance for 193 simulated trajectories in Supplementary Fig.~7a distinctly separates into two regions: one corresponding to the rotation of the terminal ethylene groups following ring-opening, and the other oscillating near 1.8~$\text{\AA}$. Consequently, trajectories exhibiting a C$_1$-C$_6$ distance greater than 3~$\text{\AA}$ at $t=$400~fs are designated as ring-opening trajectories, while those with a distance less than 3~$\text{\AA}$ are classified as ring-closing trajectories in our simulation. The criterion is consistent with previous works~\cite{Polyak2019,Medhi2024}. The corresponding evolutions of ring-opening and ring-closing trajectories are shown in Supplementary Fig.~7b, c, respectively. When the temporal resolution is taken into account in Supplementary Fig.~7d-f (i.e., by convolving the trajectories with the instrument response function), the oscillations in the ring-closing trajectories are no longer visible. This underscores the pivotal role of temporal resolution in shaping experimental observations. Therefore, extracting CI-passage dynamics requires careful consideration of temporal resolution (see Supplementary Note 8).

%%%%%%%%%%%%%%%%%%%%%%%%%%%%%%%%%%%%%
\section*{Data availability}
Datasets generated for this study and source data are available at the open-access repository~\cite{Jiang2025}.
All data supporting the conclusions are available within the article and its Supplementary Information. Source data are provided with this paper

\section*{Code availability}
Non-commercial codes used for the analysis and simulation in this work are available at the open-access repository~\cite{Jiang2025}.
%%%%%%%%%%%%%%%% REFERENCES %%%%%%%%%%%%%%%

\clearpage

\bibliography{refs} 

\bibliographystyle{naturemag}

%%%%%%%%%%%%%%%% ACKNOWLEDGEMENTS %%%%%%%%%%%%%%%

\section*{Acknowledgments}
We thank A. Natan and J. Yang for helpful discussions. This work is supported by the National Natural Science Foundation of China (Nos. 12525501, 12335010, 12450405, 11925405, 22333003 and 22361132528). D.X. would like to acknowledge the support from the New Cornerstone Science Foundation through the Xplorer Prize. H.Y. acknowledges startup funding from the School of Physical Sciences, UCSD. The UED experiment was supported by the Shanghai soft X-ray free-electron laser facility.

\section*{Author contributions}
H.J. and D.X. designed the experiments.
H.J. and T.W. tested the sample delivery system and performed the experiments.
H.J., T.W., C.J., X.Z., P.Z., T.J. and D.X. improved the UED instrument. H.J., J.Z., T.W., C.J., H.Y., Z.L., F.H. and D.X. participated in the analysis of the experimental data. H.J. performed the super-resolved real-space inversion.
J.Z., J.P. and Z.L. performed the non-adiabatic simulations. H.J., J.Z., Z.L., H.Y., F.H. and D.X. prepared the manuscript with discussion and improvements from all authors.

\section*{Competing interests}
The authors declare no competing interests.

%\clearpage

\section*{Figure}

\clearpage

\begin{center}
\section*{Supplementary Information for\\ ``\nattitle"}

	Hui Jiang$^{1,2,3\dagger}$,
	Juanjuan Zhang$^{4\dagger}$,
    Tianyu Wang$^{2,3\dagger}$,
    Jiawei Peng$^{4}$,
    Cheng Jin$^{2,3}$,\\
    Xiao zou$^{2,3}$,
    Pengfei Zhu$^{1,2}$,
    Tao Jiang$^{2,3}$,
    Zhenggang Lan$^{4\ast}$,
    Haiwang Yong$^{5\ast}$,\\
    Feng He$^{2\ast}$,
	Dao Xiang$^{1,2,3\ast}$\\
    \small$^{1}$Tsung-Dao Lee Institute, Shanghai Jiao Tong University, Shanghai 201210, China.\\
    \small$^{2}$Key Laboratory for Laser Plasmas (Ministry of Education) and School of Physics and Astronomy,\\
    \small Collaborative innovation center for IFSA (CICIFSA),\\
    \small Shanghai Jiao Tong University, Shanghai 200240, China.\\
    \small$^{3}$Zhangjiang Institute for Advanced Study, Shanghai Jiao Tong University, Shanghai 201210, China.\\
    \small$^{4}$SCNU Environmental Research Institute,\\
    \small Guangdong Provincial Key Laboratory of Chemical Pollution and Environmental Safety, \\ 
    \small MOE Key Laboratory of Environmental Theoretical Chemistry,\\
    \small School of Environment, South China Normal University, Guangzhou 510006, China.\\
    \small$^{5}$Department of Chemistry and Biochemistry, University of California San Diego, La Jolla, CA 92093, USA.\\
    \small$^\ast$Corresponding authors: zhenggang.lan@m.scnu.edu.cn (Zhenggang Lan);\\
    \small hyong@ucsd.edu (Haiwang Yong); fhe@sjtu.edu.cn (Feng He); dxiang@sjtu.edu.cn (Dao Xiang)\\
	\small$^\dagger$These authors contributed equally to this work.

\end{center}

\subsection*{This file includes:}
Supplementary Notes 1-9\\
Supplementary Figures 1 to 17\\
Supplementary References% \textit{(50-\arabic{enumiv})}\\

\newpage

%%%%%%%%%%%%%%%% MATERIALS AND METHODS %%%%%%%%%%%%%%%

\subsection*{Supplementary Note 1: Diffraction pattern processing}
Our data processing follows the below workflow.

1.~Detector mask. A mask is generated for each image. All pixels outside the phosphor screen region are masked out and the hot pixels are removed. 

2.~Background subtraction. A diffraction pattern is collected with the same integration time before delivering the sample to the gas cell.This image is then used as the background. Each loaded pattern is then subtracted from this background.

3.~Determination of the diffraction center. The center of each image is determined by fitting a ring to the pixels within a certain intensity range (typically 1500-2000 counts).

4.~Alignment of the patterns. All patterns are aligned based on the centers obtained from the fitting.

5.~Normalization. All patterns are normalized based on the total counts within a certain momentum transfer range (1-10$~\text{~\AA}^{-1}$).

6.~Average. Images at the same delay are averaged. Then, the averaged patterns for all time delays are obtained, forming a set of data.

7.~Bootstrap. A total of 300 patterns are randomly selected and averaged for each time delay to generate a new dataset. This sampling process is repeated 200 times, and the resulting 200 datasets are used to estimate the experimental uncertainty.

8.~Radial integration. The salt and pepper noise is removed by performing 2D median filtering. Then, the one-dimensional scattering intensity $I(s,t)$ is obtained by radial averaging around the diffraction center.

The above procedures delineate the method for obtaining the 1D $I(s,t)$ from the 2D measured diffraction pattern. The static scattering intensity $I(s,t<0)$ comprises the molecular scattering signal $I_m(s,t<0)$, atomic scattering signal $I_a(s,t<0)$ and other background $I_b(s,t<0)$. To extract  $I_m(s,t<0)$, we follow the methods used by Ihee~\cite{Ihee2002}. First, zero points of $I_m^{\text{sim}}(s,t<0)$ from theoretical results are calculated. Then, we fit the values of $I(s,t<0)$ at these zero points with a low-order polynomial. The fitted curve contains both the background and atomic scattering signals. Finally, the static molecular scattering intensity $I_m(s,t<0)$ is obtained by subtracting the fitted curve from $I(s,t<0)$. The modified scattering intensity (sM) is computed by the following equation:
\begin{equation}
    {\rm{sM}}(s)=\frac{I_m(s,t<0)}{I_a}s,
\end{equation}
The atomic scattering intensity is calculated as follows: 
\begin{equation}
    I_a=\sum_{i}^{N}f_i^*(s)f_i(s)
\end{equation}
where $f_i(s)$ is the elastic scattering amplitude for the $i$th atom and is calculated by the ELSEPA program~\cite{Salvat2005}.

The difference scattering intensity is computed as follows:
\begin{equation}
    {\Delta{I}}(s,t)=I(s,t)-I(s,t<0).
\end{equation}
Both the background and the atomic signal are subtracted without further processing.
Then the modified difference scattering intensity can be calculated as follows:
\begin{equation}
    \Delta {\rm{sM}}(s,t)=\frac{{\Delta{I}}(s,t)}{I_a}s.
    \label{cal_deltasM}
\end{equation}
Due to the relatively low signal-to-noise ratio in the high-$s$ region, a baseline modification to $\Delta {\rm{sM}}(s>5.5~\text{\AA})$ by fitting a low-order polynomial to the signal is used, similar to the method used by other groups~\cite{Yang2020,Ihee2002}. The $\Delta {\rm{sM}}(s,t)$ before and after the correction are shown in Supplementary Fig.~\ref{sup3}.

\subsection*{Supplementary Note 2: Linear absorption range}
In order to minimize the impact of multiphoton absorption and ionization on the experimental results, it is necessary to carefully select the pulse energy for the UV pump laser. To this end, the integrated absolute intensity of  $I(t=2~{\rm{ps}})-I(t=-1~{\rm{ps}})$ at $1.5<s<6~\text{\AA}^{-1}$ was measured under varying laser intensities. The results are shown in Supplementary Fig.~\ref{suplinear}. The intensity exhibited a linear relationship with the pulse energy. In this study, the pulse energy was maintained at approximately 50~$\rm{\mu J}$, which is within the linear region.

\subsection*{Supplementary Note 3: Details in pair distribution function (PDF) transform}
In the measured static sM($s$), the signals at $s<0.8$~$\text{\AA}^{-1}$ are not measured due to the central hole of the phosphor screen. Additionally, the procedure of removing the background by polynomial fitting mentioned above cannot accurately eliminate the background signal at small scattering angles, leading to deviations between the experimental and theoretical values at $0.8<s<1.3$~$\text{\AA}^{-1}$. This Unremoved background signal may be attributed to the unscattered electron beam. Noted that the background signals can be naturally eliminated when calculating the difference signals. To circumvent the occurrence of non-physical outcomes stemming from the truncation of the low-$s$ region when computing the static PDF, scaled simulation results are used to supplement the data within the range of $0<s<1.3$~$\text{\AA}^{-1}$.

For the modified difference scattering intensity $\Delta$sM($s,t$), although the background has been subtracted when computing $\Delta I$, the inelastic scattering signals in the low-$s$ region ($s<1.3~\text{\AA}^{-1}$) can still affect the structural information. Thus, we fill $\Delta$sM($s,t$) in the range of 0-1.3~$\text{\AA}^{-1}$ by extrapolation of the experimental data. We fit the $\Delta$sM($s$) at $1.3<s<4~\text{\AA}^{-1}$ with the model:
\begin{equation}
    {\rm{sM}}_{\rm{extra}}(s)=\sum_{k=1}^{N}\alpha_k\frac{\sin(sr_k)}{sr_k}s,
\end{equation}
in which $\alpha_k$ and $r_k$ are the optimization parameters. In consideration of the asymptotic behavior of 
$\Delta$sM($s$) at $s=0$, we impose the constraint $\sum_{k=1}^{N}\alpha_k=0$ and $N$ is set to 4. Subsequently, the $\Delta$sM($s$) signals at $0<s<1.3~\text{\AA}^{-1}$ are filled by the fitting function. The complete $\Delta$sM($s,t$) after extrapolation is shown in Supplementary Fig.~\ref{sup4} where the inelastic scattering signals have been removed. We compare the $\Delta {\rm{PDF}}(r,t=1~\rm{ps})$ calculated by two different integration intervals ($1.3<s<10~\text{\AA}^{-1}$ and $0<s<10~\text{\AA}^{-1}$), as shown in Supplementary Fig.~\ref{sup5}. It can be seen that filling the signals at $s<1.3~\text{\AA}^{-1}$ only results in a slight change in the intensity of certain peaks without affecting the overall characteristics of $\Delta {\rm{PDF}}$.

It is important to note that all the methods described above for filling sM at low-$s$ region are not required in the super-resolved inversion technique.

\subsection*{Supplementary Note 4: Super-resolved pair distribution function (s-PDF) transform}
Here, we provide a concise overview of the procedures employed to achieve super-resolution ~\cite{Natan2023} in this experiment.

1.~ Discretization in momentum and real space. A schemed discretization is applied in the real-space inversion:
\begin{equation}
    \begin{cases}
        s_{i}=\frac{j_{0i}}{j_{0N}}N\Delta s,~~~&1<i<N \\
        R_{m}=\frac{j_{0m}}{M\Delta s},~~~&1<m<M  
    \end{cases}
\end{equation}
in which $j_{0i}$ is the $i$th root of $j_0$ (the zeroth-order spherical Bessel function of the first kind). In this study, the discrete spacing in momentum space, denoted by $\Delta s$, is set to 0.1$~\text{\AA}^{-1}$. Given the maximum effective $s$-range ($\sim10~\text{\AA}^{-1}$), N is set to 101 and therefore $s_{max}=N\Delta s= 10.1~\text{\AA}^{-1}$. M=ceil($\pi/(\Delta R\Delta s)$)+1. $\Delta R$ is chosen to be 0.05~$\text{\AA}$, and the corresponding super-resolution factor is $M/N\approx6$.

2.~Window function. The window function $h(s)$ is chosen to be the same as that for computing PDF:
\begin{equation}
   h(s)=
    \begin{cases}
        e^{-ks^2},~~~&s_{min}<s<s_{max} \\
        0,~~~&s<s_{min} 
    \end{cases}
\end{equation}
Given the effective $s$-range, $k$ is set to 0.03~$\text{\AA}^2$.  In the super-resolution inversion approach, we take $s_{min}=1.3~\text{\AA}^{-1}$ and therefore we no longer need to fill the scattering signals in the low-$s$ region ($s<1.3~\text{\AA}^{-1}$), as was done in the PDF transform.

3.~Dictionary of natural scattering kernel (NSK). Super-resolution reconstructs the original pair distance density ($\rho(R)$) with a series of weighted $\delta$-function pair distance density. NSK corresponds to the real-space transformation of individual $\delta$-function density ($\rho(R)=\delta(R-R_{m^*})$):
\begin{equation}
    {\rm{NSK}}_{m^*}(R_m)=\frac{[(M-1)\Delta s]^2}{j_{0M}}\sum_{i=1}^{M-1}G_{mi}j_0(s_iR_{m^*})s_ih(s_i),
   \label{NSK}
\end{equation}
\begin{equation}
    G_{mi}=2\frac{j_0(\frac{j_{0m}j_{0i}}{j_{0M}})}{j_{0M}j_1^2(j_{0i})}.
\end{equation}
Then the dictionary is composed of a set of NSKs along the real-space sampling $R_{m^*}$:
\begin{equation}
    \mathcal{D}=
    \begin{bmatrix}
    \mathrm{NSK}_{1^{*}}(R_1)&\cdots & \mathrm{NSK}_{m^{*}}(R_1) & \cdots& \mathrm{NSK}_{M^{*}}(R_1)\\
    \vdots & & \vdots & & \vdots \\
    
    \mathrm{NSK}_{1^{*}}(R_m) & \cdots & \mathrm{NSK}_{m^{*}}(R_m) & \cdots & \mathrm{NSK}_{M^{*}}(R_m) \\
    
    \vdots & & \vdots & & \vdots\\
    
     \mathrm{NSK}_{1^{*}}(R_M)& \cdots& \mathrm{NSK}_{m^{*}}(R_M) & \cdots& \mathrm{NSK}_{M^{*}}(R_M)
    
    \end{bmatrix}.
\end{equation}

4.~ Inversion of scattering signals. The discretized scattering signals are converted to the pair distribution function in real space by:
\begin{equation}
    \Delta{{\rm{PDF}}}(R_m,t)=\frac{[(M-1)\Delta s]^2}{j_{0M}}\sum_{i=1}^{M-1}G_{mi}\frac{\Delta I(s_i,t)}{f_e(s_i)}s_ih(s_i),
    \label{deltapd}
\end{equation}
where $\Delta I(s,t)$ is the diffraction signals that contains the structural information. In the case of static structures, $\Delta I(s,t)$ corresponds to the molecular diffraction signal. Similarly, for time-resolved structural changes, it refers to the difference diffraction signals. $f_e(s)$ is the effective atomic scattering factor and is estimated by $f_e(s)=f_a^{*}(s)f_b(s)$, where $a$ and $b$ refer to the main types of atoms. In this study, C-C pair distances dominate, and as a result, $f_e(s)$ is set to $f_C^{*}(s)f_C(s)$.

5.~Deconvolution of $\Delta{\rm{PDF}}$. The dictionary generated above is used to deconvolute $\Delta \rm{PDF}$ by minimizing:
\begin{equation}
    ||\Delta {\rm{PDF}}-\mathcal{D}\boldsymbol{w}||^2 +\alpha \mathcal{R}(\boldsymbol{w}),
\end{equation}
where $\mathcal{R}(\boldsymbol{w})=\sum_{m}|\boldsymbol{w}_m|$. This $l_1$ regularizer~\cite{Tibshirani1996} gives us sparse weighted $\delta$-function pair density with $\boldsymbol{w}_m$ being the weight. $\alpha$ is determined by the corner of L-curve~\cite{Hansen1993}. The convex optimization is performed with a package~\cite{Convexjl2014}. For static s-PDF, we impose the constraint $\boldsymbol{w}_m\geq0$. 

To understand the meaning of the intensity of s-PDF, we consider a simple case when the molecule only consists of one type of atom. In this case, Eq.~(\ref{deltapd}) can be further simplified to:
\begin{equation}
        \Delta{\rm{PDF}}(R_m,t)=\frac{[(M-1)\Delta s]^2}{j_{0M}}\sum_{i=1}^{M-1}\sum_{R} G_{mi}\Delta \rho(R,t) j_0(s_iR)s_ih(s_i),
        \label{deltapd_1}
\end{equation}
where $\Delta \rho(R,t)$ refers to the difference pair charge density. Comparing Eq.~(\ref{deltapd_1}) with the NSK~(\ref{NSK}), we can see that $\boldsymbol{w}(m)\approx\Delta \rho(R=R_m,t)$ if R is sparse. Therefore, the super-resolved intensity directly reflects the pair density for molecules comprising solely one type of atom. For molecules comprising different types of atoms, the super-resolved intensity is proportional to the pair density and the product of scattering cross sections of the atom pairs.

The validity of the deconvolution can be further substantiated by employing the L-curve method depicted in Supplementary Fig.~\ref{sup6}. The L-shaped curves in both the static s-PDF and $\Delta$s-PDF justify the choice of $l_1$ regularizer.

\subsection*{Supplementary Note 5: Static C-H pair distribution}
The static scattering signals encode the ground-state CHD structural information with a high signal-to-noise ratio. The super-resolution technique facilitates the reconstruction of diverse interatomic distances, including all C-C and C-H interatomic distances for static structures. As previously mentioned, the super-resolved peak intensities are proportional to the product of the pair density and the scattering cross sections of the atom pairs. To facilitate comparison with s-PDF, here the distribution of the interatomic distance in the static structures obtained through theoretical sampling is scaled with the scattering cross sections of the atomic pairs and shown in Supplementary Fig.~\ref{supstatic}. To gain more insights, we separate the different types of atomic pairs. As shown in Supplementary Fig.~\ref{supstatic}b, the C-C interatomic distances exhibit three distinct peaks corresponding to $R_1$, $R_2$ and $R_3$. The C-H interatomic distances in Supplementary Fig.~\ref{supstatic}c exhibit peaks at 1.05, 2.1, 3.5, and 3.9~$\text{\AA}$, corresponding to the 1st, 2nd, 3rd, and 4th C-H interatomic distances, respectively. The small peak near 2.9 $\text{\AA}$ overlaps with the $R_3$ C-C pair distance. The weight of the H-H interatomic distances in Supplementary Fig.~\ref{supstatic}d is too small to be clearly observed. Nevertheless, the reconstruction of characteristic C-H pair distances has been accomplished with high fidelity, underscoring the distinct advantage of s-PDF in achieving high spatial resolution.

\subsection*{Supplementary Note 6: Background subtraction of the inelastic scattering signal}
The intensity of the inelastic scattering signal is obtained by integrating over the $0.8<s<1.05~\text{\AA}^{-1}$ region in $\Delta$sM, as shown in Supplementary Fig.~\ref{sup_inelastic}a. While the signal's primary origin is inelastic scattering, it also contains contributions from elastic scattering and the residual background. Consequently, we subtract the extrapolated elastic scattering signals and the remaining constant background signals which may be induced by the ionization effects, and the modified inelastic scattering signals are shown in Supplementary Fig.~\ref{sup_inelastic}b. It is evident that these procedures exert minimal influence on the rising edge of the signal, with only a small effect on the fitted decay time constant at the falling edge.

\subsection*{Supplementary Note 7: The influence of different atomic pairs on the interpretation of CI dynamics.}
Since the scattering signals are dominated by carbon atoms, we focus on the influence of C-C pairs on structural dynamics. In addition to the analyzed C$_1$-C$_6$ pair, Supplementary Figure~\ref{sup_c1c2_c1c3} shows the evolution of C$_1$-C$_2$ (nearest) and C$_1$-C$_3$ (next-nearest) obtained using the simulated trajectories. The dynamics of these two interatomic distances can help us assess the impact of molecular distortions beyond the contributions from the C$_1$-C$_6$ pair and explain why the evolution of C$_1$-C$_6$ was used to qualitatively analyze the characteristic changes in atomic pair density near 1.95~$\text{\AA}$ and 2.3~$\text{\AA}$ within the first 100~fs, respectively.

We first consider the $\Delta$s-PDF near 1.95~$\text{\AA}$. As shown in Supplementary Fig.~\ref{sup_c1c2_c1c3}a, in the absence of C$_1$-C$_6$ pair, the nearest atomic pairs arising from molecular distortion or vibration typically contribute to signals below 1.75~$\text{\AA}$, and thus have minimal impact at 1.95~$\text{\AA}$. Also, Supplementary Figure~\ref{sup_c1c2_c1c3}b shows that next-nearest pairs generally influence regions beyond 2.1~$\text{\AA}$, indicating that their contribution to the 1.95~$\text{\AA}$ signal is also negligible. Therefore, the feature at 1.95~$\text{\AA}$ primarily originates from the early-stage opening of the C$_1$-C$_6$ pair, as it approaches and crosses the 1.95~$\text{\AA}$ region.

Since the excess kinetic energy of the molecule reaches its maximum after the electronic state decays to its ground state, structural distortions and vibrational motions are not pronounced during the early stage of the CI passage. As shown in Supplementary Fig.~\ref{sup_c1c2_c1c3}b, d, the evolution of the next-nearest pair remains largely unchanged within the first 80~fs. Significant C$_1$-C$_3$ oscillations only appear after the wave packet passes through the second CI, when the molecule gains substantial kinetic energy. Therefore, although the $\Delta$s-PDF feature near 2.3~$\text{\AA}$ lies within the typical range of R$_2$, during the initial 80~fs of CI dynamics, the signal arising from molecular distortions is relatively weak. The observed knee feature at $\sim$2.3~$\text{\AA}$ in the $\Delta$s-PDF is primarily attributed to the progressive elongation of the C$_1$-C$_6$ pair distance approaching this distance.

For these reasons, we use the evolution of the C$_1$-C$_6$ pair to qualitatively illustrate the structural dynamics near the conical intersections.

\subsection*{Supplementary Note 8: Effect of temporal resolution on imaging the CI dynamics}
In order to facilitate a reliable comparison between experiment and theory, all the simulated results in this study have been convoluted with the instrument response function (IRF) of the measurement. The s-PDF analysis reveals that the time for the wave packet to traverse through the two CIs is approximately 30~fs, which is smaller than the IRF (about 80~fs). While resolving a time-delay smaller than the IRF of the measurement is not uncommon, this analysis is only feasible when the temporal resolution is sufficiently high. To illustrate the effect of temporal resolution on the extracted information of the CI dynamics, the simulated intensity of the $\Delta$s-PDF at $\sim$1.95$\text{~\AA}$ and $\sim$2.30$\text{~\AA}$ with various IRF is shown in Supplementary Fig.~\ref{sup_temp}. The results indicate that when the IRF is larger than 120~fs, the characteristic features employed for CI dynamics extraction vanish. In this study, the 80-fs IRF provides the sufficient resolution to resolve both the characteristic structures of $\Delta$s-PDF at $\sim$1.95$\text{~\AA}$ and $\sim$2.30$\text{~\AA}$, which are crucial for extracting the traversal time between the two CIs.

\subsection*{Supplementary Note 9: Effect of $s$-range on imaging the CI dynamics}
To simultaneously minimize the influence of inelastic scattering and enhance spatial resolution, we chose a relatively small lower limit $s=1.3~\text{\AA}^{-1}$. To validate whether different choices of the $s$-range affect the observed CI structural dynamics, here we show the structural dynamics during passage through the two CIs calculated by two different $s$-ranges in Supplementary Fig.~\ref{sup_srange}. The lower limit of the momentum transfer used for the $\Delta$s-PDF transformation is set to 1.5~$\text{\AA}^{-1}$ in Supplementary Fig.~\ref{sup_srange}a, c and 1.7~$\text{\AA}^{-1}$ in Supplementary Fig.~\ref{sup_srange}b, d. By comparing with the results in Fig. 4a, c of the main text, we find that varying the lower limit of the $s$-range has no noticeable effect on the characteristic features observed in the $\Delta$s-PDF evolution (knee structure). These findings indicate that our results are robust against different $s$-ranges. Qualitatively, this robustness arises because the dynamics near the conical intersection involve relatively small structural changes, which primarily contribute to scattering signals at larger scattering angles.

%%%%%%%%%%%%%%%%%%%%%%%reference%%%%%%%%%%%%%%%%%%%%%%%%%%

%\renewcommand{\refname}{Supplementary References}

%\bibliographystyle{naturemag}
%\bibliography{refs}

%%%%%%%%%%%%%%%% SUPPLEMENTARY FIGURES %%%%%%%%%%%%%%%

\makeatletter
\renewcommand{\fnum@figure}{\textbf{Supplementary Fig. \thefigure}}
\renewcommand{\fnum@table}{\textbf{Supplementary Table \thetable}}
\makeatother
\setcounter{figure}{0}
\setcounter{table}{0}

\clearpage

\begin{figure} 
	\centering
	\includegraphics[width=1\textwidth]{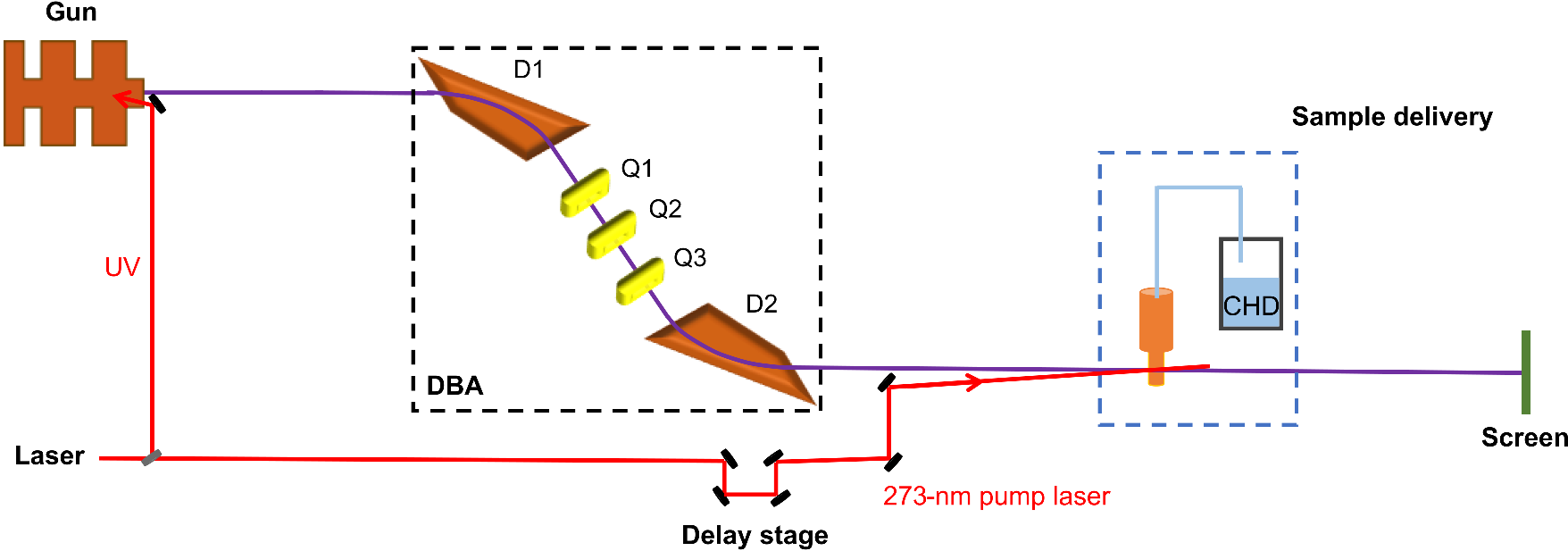} 
	\caption{\textbf{Experimental setup.}
	The focused 273-nm laser pulse intersects with the electron beam on the sample delivered via a flow cell. The electron beam is compressed by a double bend achromat (DBA) which consists of two dipole magnets (D1 and D2) and three quadrupole magnets (Q1, Q2 and Q3) to reduce both the bunch length and arrival time jitter. The instrument response function in this experiment was $\sim$80~fs (FWHM).}
	\label{ext_setup}
\end{figure}

\begin{figure}
	\centering
	\includegraphics[width=0.6\textwidth]{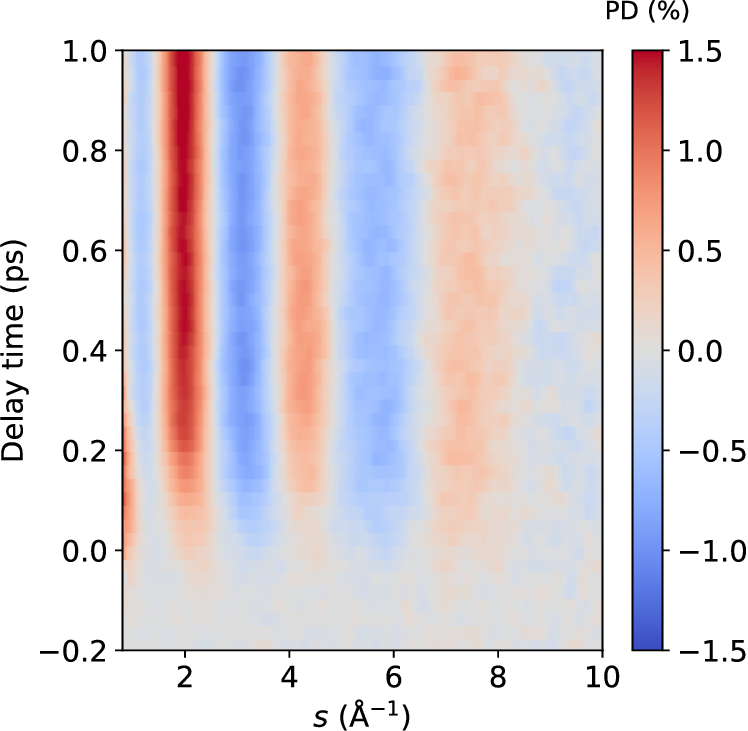} 
	\caption{\textbf{The full data set of experimental percentage difference signal.}Source data are provided as a Source Data file.
		}
	\label{ext_PD}
\end{figure}

\begin{figure} 
	\centering
	\includegraphics[width=0.9\textwidth]{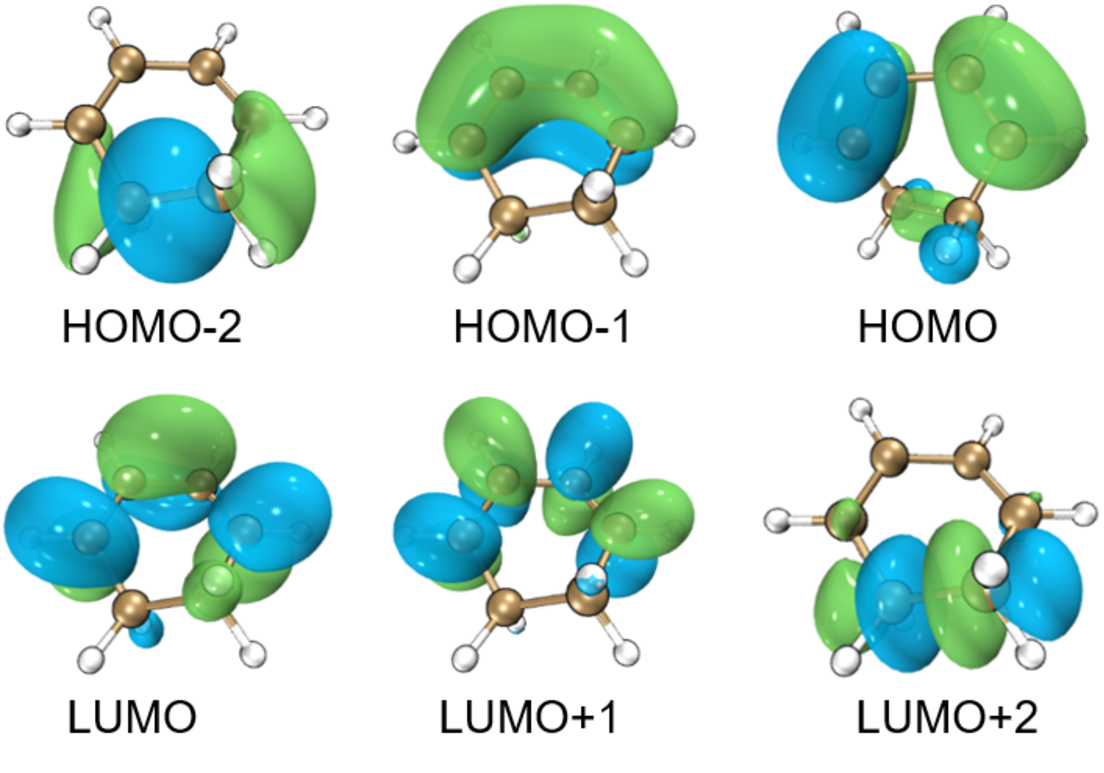} 
	\caption{\textbf{Active space orbitals in the XMS (3)-CASPT2(6, 6)/def2-TZVP calculations.}
		}
	\label{ext_activespace} 
\end{figure}

\begin{figure} 
	\centering
	\includegraphics[width=0.9\textwidth]{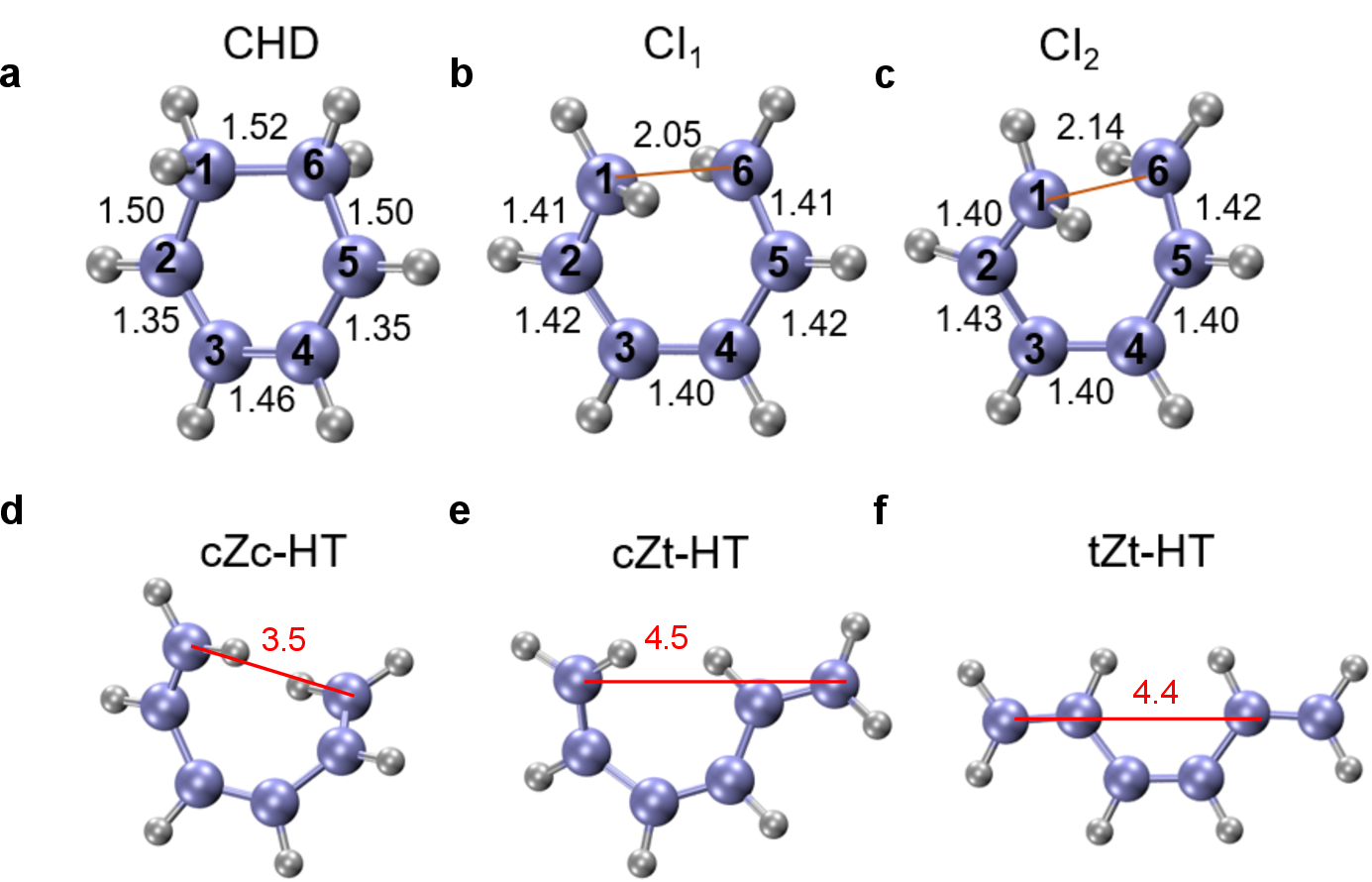} 
	\caption{\textbf{The optimized geometries of S$_0$-min, CI$_1$, CI$_2$ and three HT isomers.}
		\textbf{a}, S$_0$-min. \textbf{b}, CI$_1$. \textbf{c}, CI$_2$. \textbf{d}, cZc-HT. \textbf{e}, cZt-HT. \textbf{f}, tZt-HT. Interatomic distances are given in units of $\text{\AA}$. Source data are provided as a Source Data file.}
	\label{ext_geometries} 
\end{figure}

\begin{figure} 
	\centering
	\includegraphics[width=0.6\textwidth]{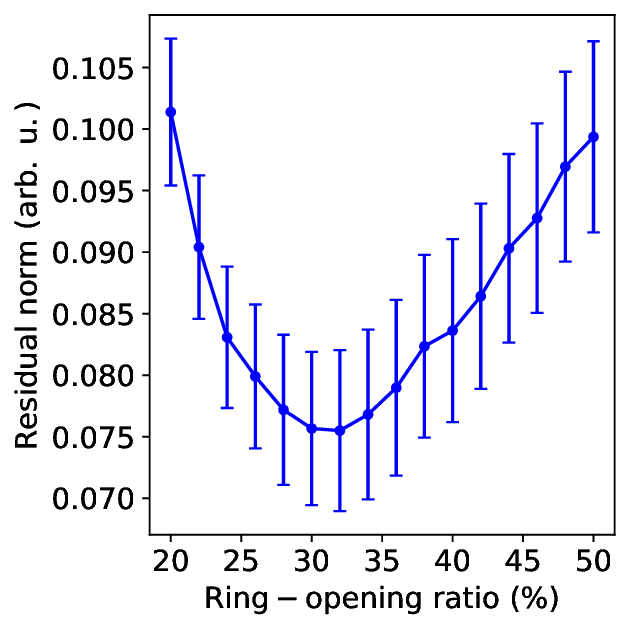} 
	\caption{\textbf{Estimation of the quantum yield of HT.}
    The residual norm under different ring-opening ratios is defined as the difference between the simulated and measured values of the $\Delta$sM over the same time interval, with the simulated values calculated using different ring-opening ratios. Error bars represent a 68\% confidence interval. Source data are provided as a Source Data file.
		}
	\label{ext_openrate} 
\end{figure}

\begin{figure} 
	\centering
	\includegraphics[width=0.6\textwidth]{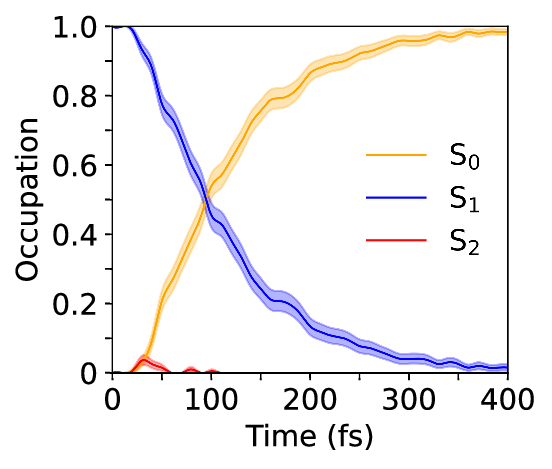} 
	\caption{\textbf{The population evolutions of excited states as a function of delay time in the nonadiabatic simulation.}
    The uncertainties of these curves are given by the Bootstrap method.}
	\label{ext_population} 
\end{figure}

\begin{figure} 
	\centering
	\includegraphics[width=0.9\textwidth]{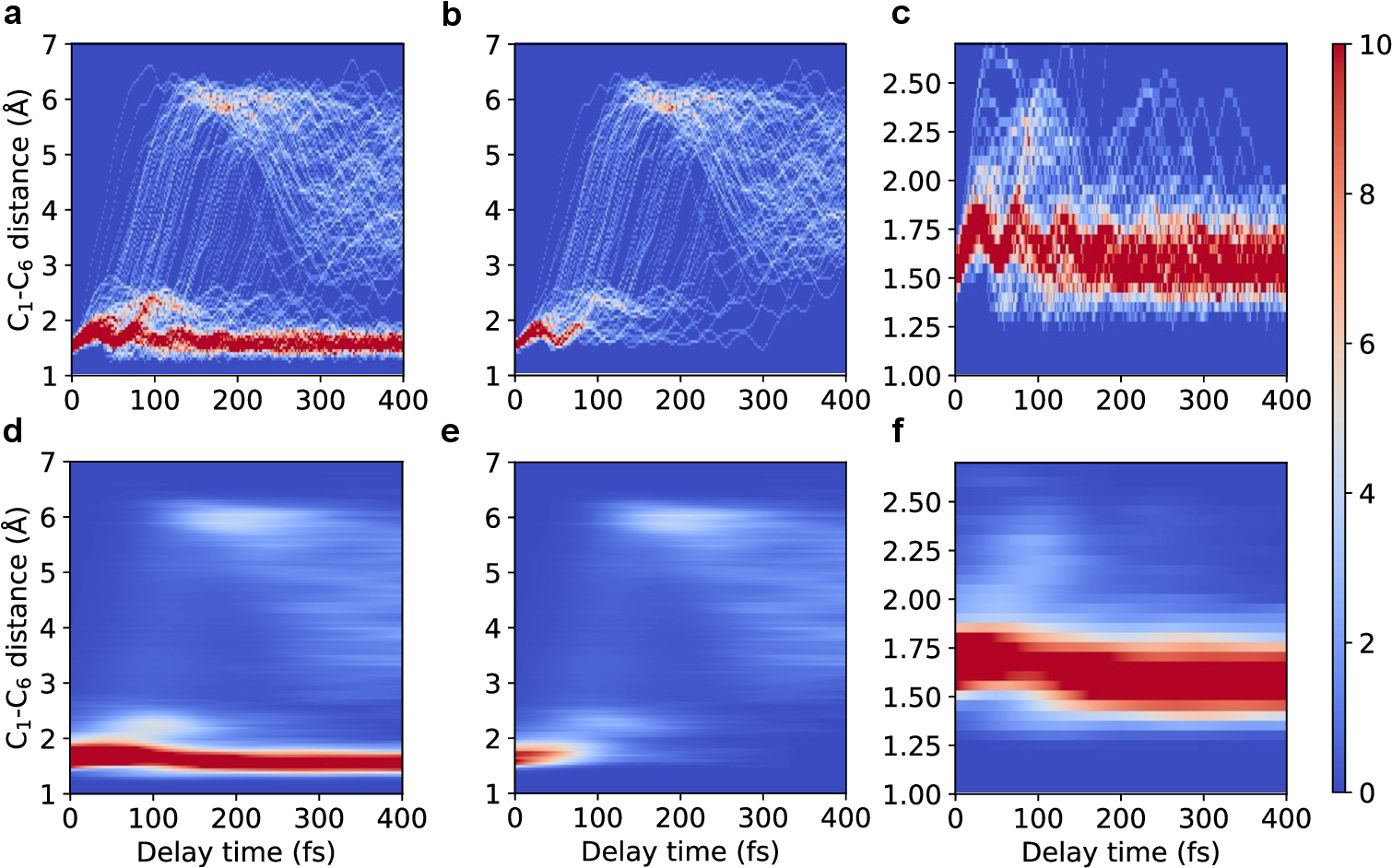} 
	\caption{\textbf{Simulated evolutions of the C$_1$-C$_6$ interatomic distance.}
    \textbf{a}, All 193 trajectories. \textbf{b}, Ring-opening trajectories. \textbf{c}, Ring-closing trajectories. \textbf{d}-\textbf{f}, The same sets of trajectories shown in panels \textbf{a}-\textbf{c}, respectively, after convolution with a Gaussian function (80-fs FWHM) to account for temporal resolution. Source data are provided as a Source Data file.}
	\label{ext_c1c6} 
\end{figure}

\begin{figure} 
	\centering
	\includegraphics[width=0.9\textwidth]{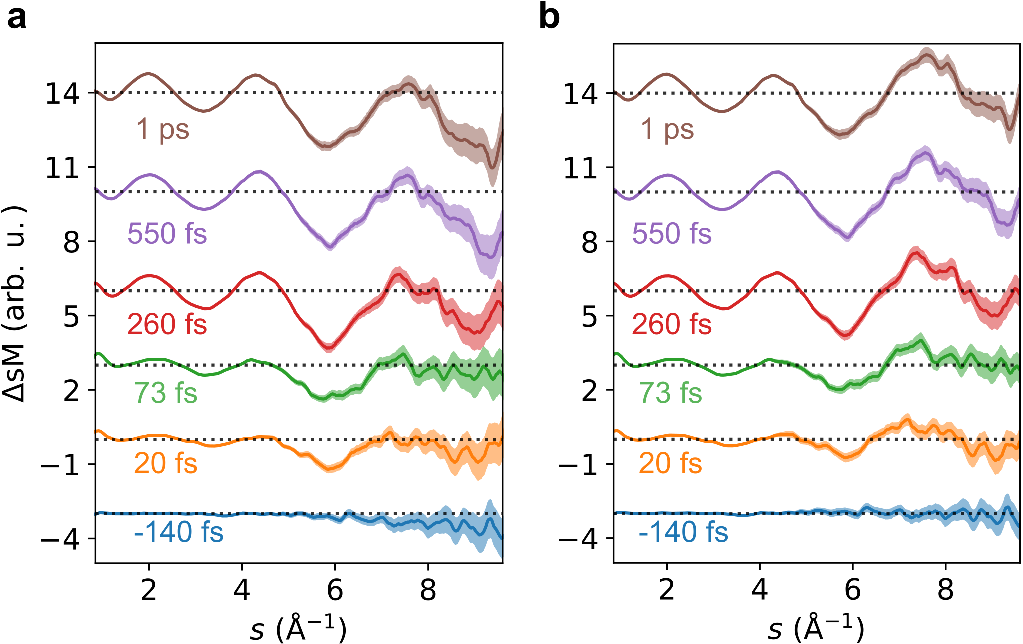} 
	\caption{\textbf{Baseline modification for high-$s$ signals.}
		\textbf{a} and \textbf{b}, The $\Delta$sMs at six selected time points before and after the baseline modification by a low-order polynomial fit performed on $\Delta$sM in the range of $s>5.5~\text{\AA}^{-1}$, respectively. The black dotted lines are the baselines for each curve. The uncertainties are calculated by the bootstrap method and the shaded regions denote the 68\% confidence interval. Source data are provided as a Source Data file.}
	\label{sup3} 
\end{figure}

\begin{figure} 
	\centering
	\includegraphics[width=0.6\textwidth]{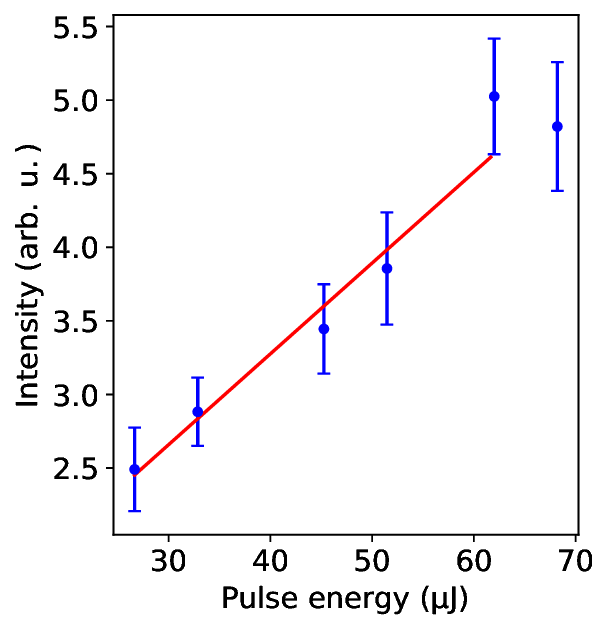} 
	\caption{\textbf{The pump pulse energy scan.}
		Correlation between the integrated absolute diffraction intensity change in the range of $1.5<s<6~\text{\AA}^{-1}$ and the UV pulse energy. Error bars represent a 68\% confidence interval. Source data are provided as a Source Data file.}
	\label{suplinear} 
\end{figure}

\begin{figure} 
	\centering
	\includegraphics[width=0.6\textwidth]{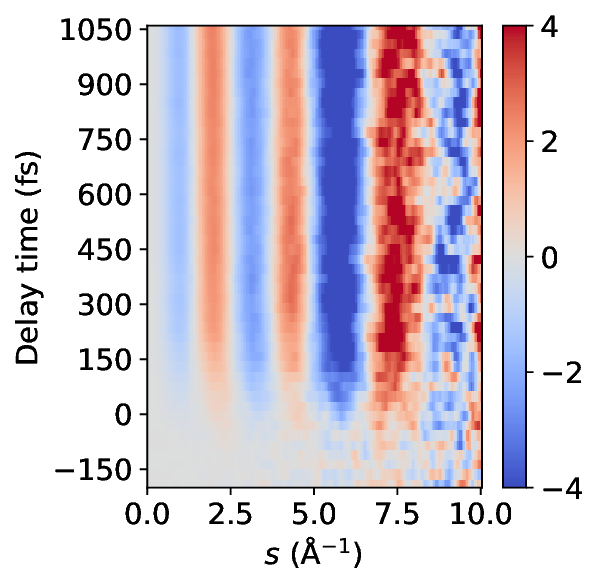} 
	\caption{\textbf{Extrapolation of elastic scattering signals. }
		The measured elastic scattering $\Delta$sM involves the removal of the original signals at $s<1.3~\text{\AA}^{-1}$, which contain inelastic scattering, and the supplementation of these signals with extrapolations derived from experimental data within the range of $1.3<s<4~\text{\AA}^{-1}$. Source data are provided as a Source Data file.}
	\label{sup4} 
\end{figure}

\begin{figure} 
	\centering
	\includegraphics[width=0.6\textwidth]{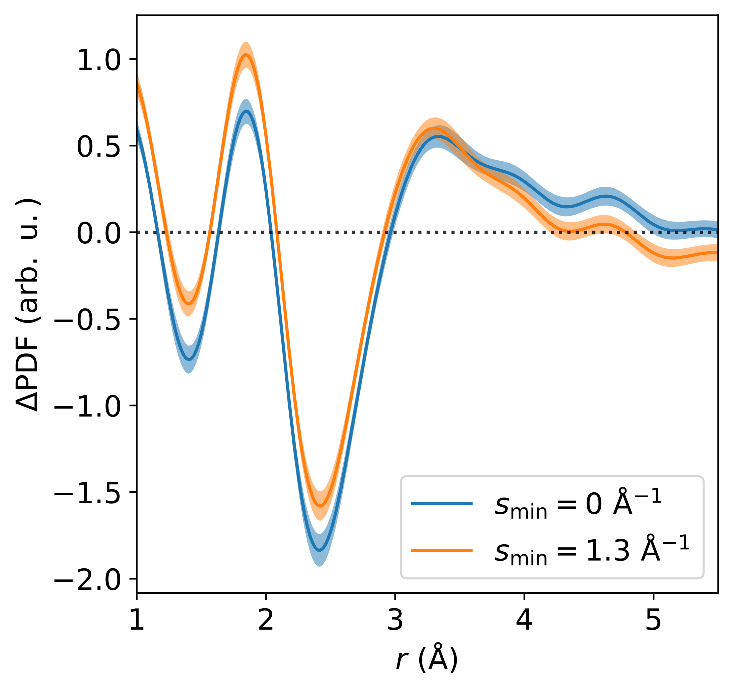} 
	\caption{\textbf{$\Delta$PDFs computed with different $s$-range. }
		The orange and blue lines represent the $\Delta$PDFs before and after the extrapolation of low-$s$ elastic scattering signals, respectively. The shaded regions denote the 68\% confidence interval. Source data are provided as a Source Data file.}
	\label{sup5} 
\end{figure}

\begin{figure} 
	\centering
	\includegraphics[width=0.6\textwidth]{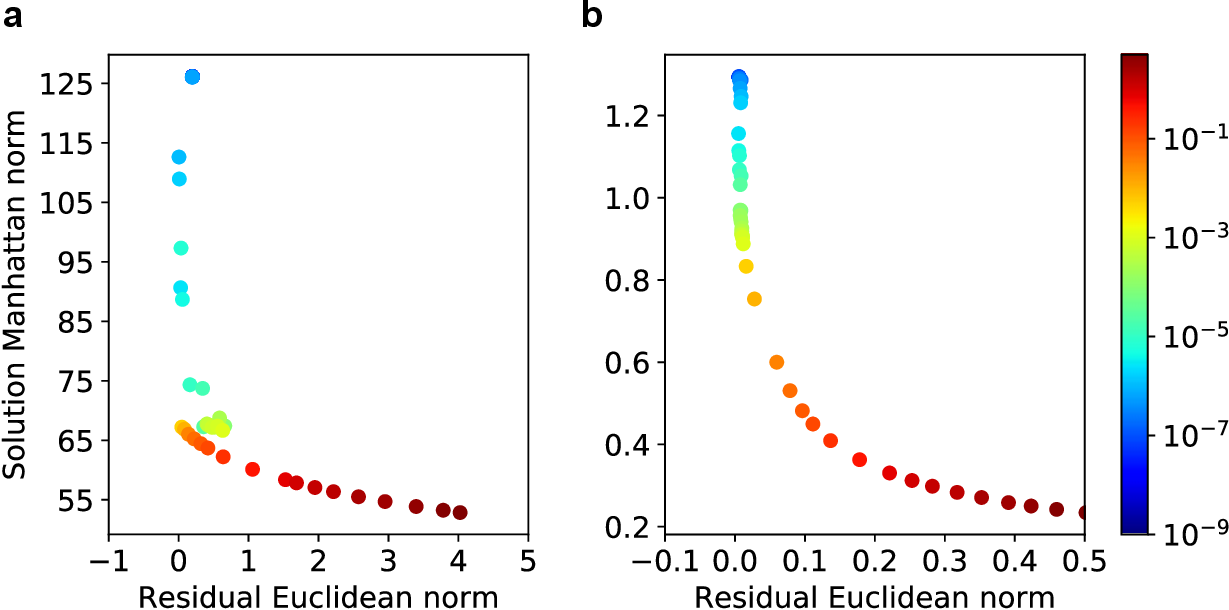} 
	\caption{\textbf{The L-curve representations in optimizations of static s-PDF and $\Delta$s-PDF.}
		The L-curves, which plot the trade-off between the residual norm and solution Manhattan norm when computing static s-PDF in \textbf{a} and $\Delta$s-PDF in \textbf{b}, are used to determine the regularization parameters and verify the choice of the regularization. The regularization parameter $\alpha$ is set near the corner of the L-shaped curve. Source data are provided as a Source Data file.}
	\label{sup6} 
\end{figure}

\begin{figure} 
	\centering
	\includegraphics[width=0.6\textwidth]{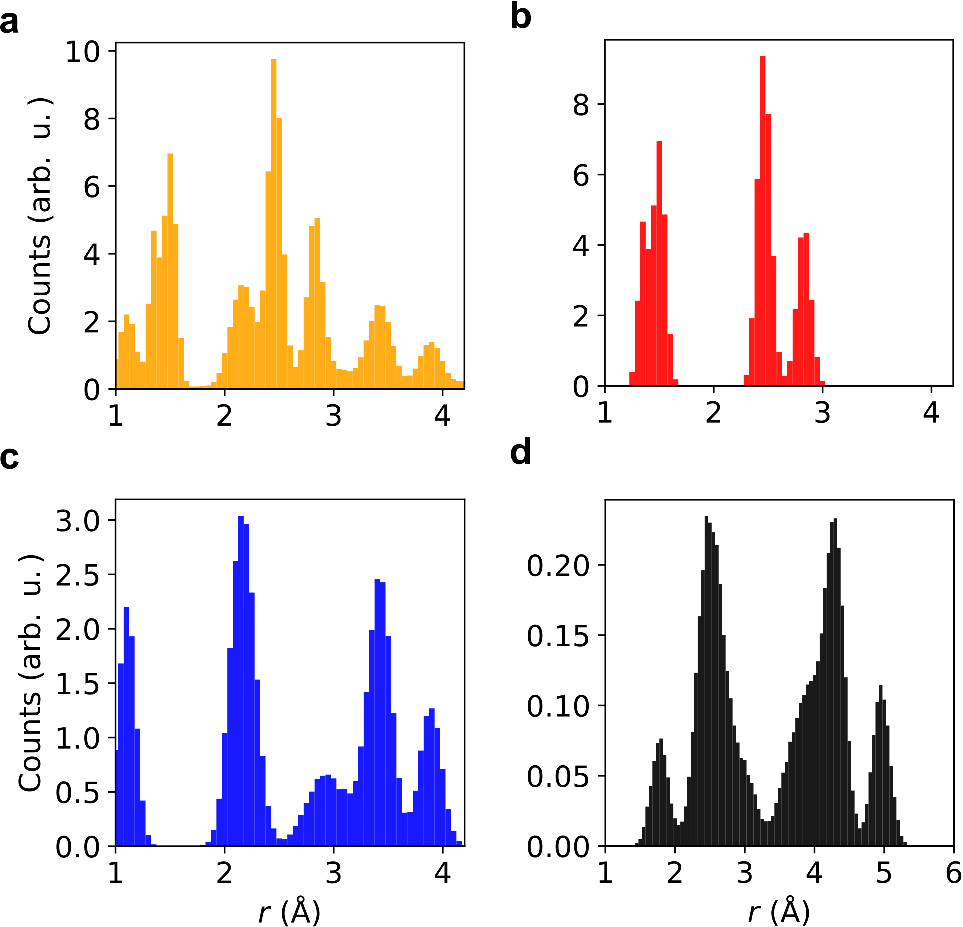} 
	\caption{\textbf{The pair distance distribution of static simulated structures.}
    \textbf{a}, All pair distances. \textbf{b}, C-C pair distances. \textbf{c}, C-H pair distances. \textbf{d}, H-H pair distances. A total of 5,000 conditions sampled from the Wigner distribution function of the lowest vibrational state at the S$_0$-min are counted. The count of each pair distance is scaled by the product of their atomic scattering amplitudes. Source data are provided as a Source Data file.
	}
	\label{supstatic} 
\end{figure}

\begin{figure} 
	\centering
	\includegraphics[width=0.8\textwidth]{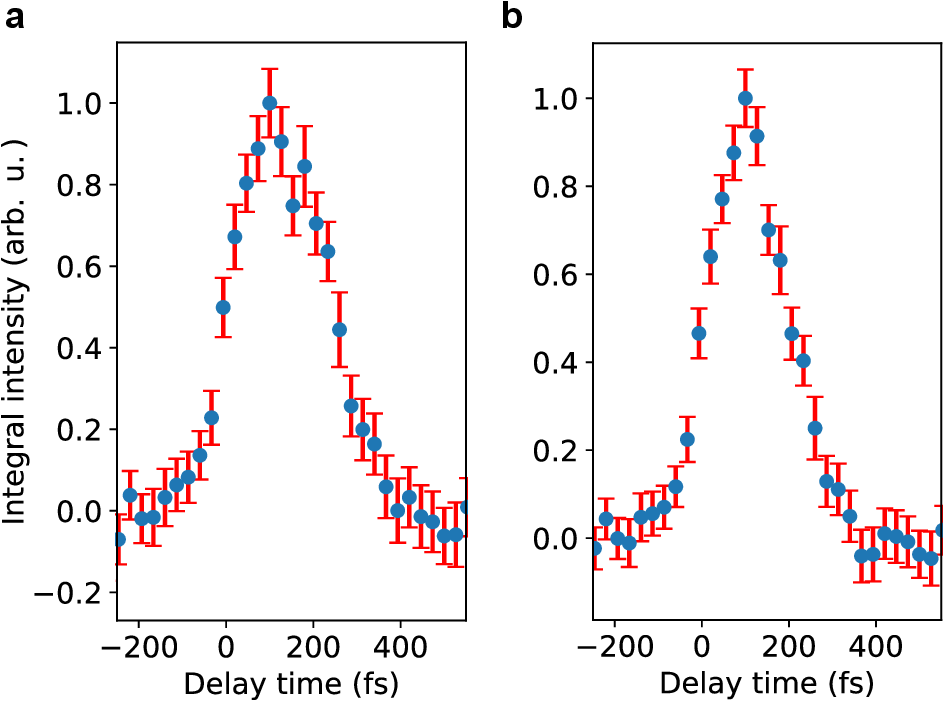} 
	\caption{\textbf{Background removal of the inelastic scattering signal.}
    \textbf{a}, The original integrated intensity of small-angle signals at $0.8<s<1.05~\text{\AA}^{-1}$.
    \textbf{b}, The modified integrated intensity after removing the elastic signals and the additional background.
    Error bars represent a 68\% confidence interval. Source data are provided as a Source Data file.
		}
	\label{sup_inelastic} 
\end{figure}

\begin{figure} 
	\centering
	\includegraphics[width=0.8\textwidth]{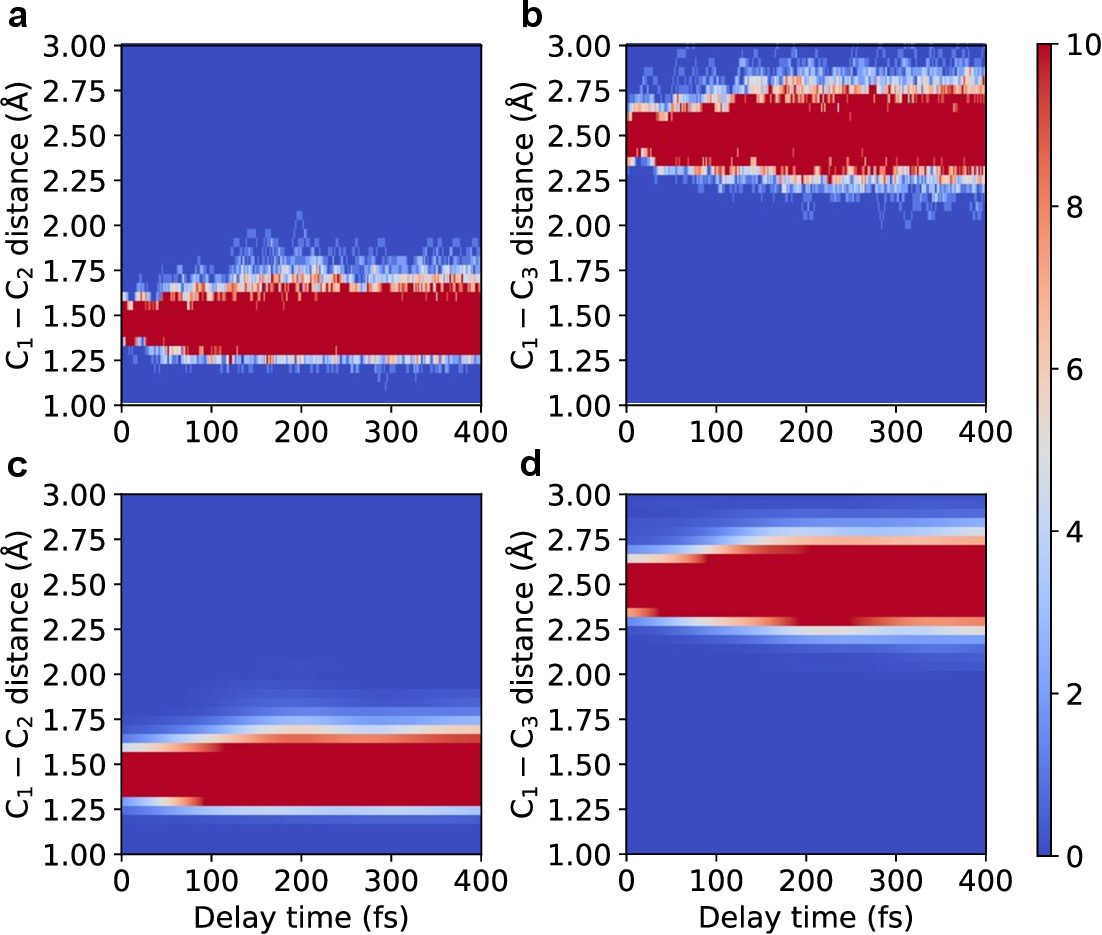} 
	\caption{\textbf{Simulated evolutions of C$_1$-C$_2$ and C$_1$-C$_3$ interatomic distances from all 193 trajectories.} Atomic labels are shown in Fig. 1b. C$_1$-C$_2$ refers to a nearest carbon pair, while C$_1$-C$_3$ corresponds to a next-nearest pair. \textbf{a} and \textbf{c}, The evolution of C$_1$-C$_2$ before and after temporal convolution. \textbf{b} and \textbf{d}, The evolution of C$_1$-C$_3$ before and after temporal convolution. Source data are provided as a Source Data file. }
	\label{sup_c1c2_c1c3} 
\end{figure}

\begin{figure} 
	\centering
	\includegraphics[width=0.8\textwidth]{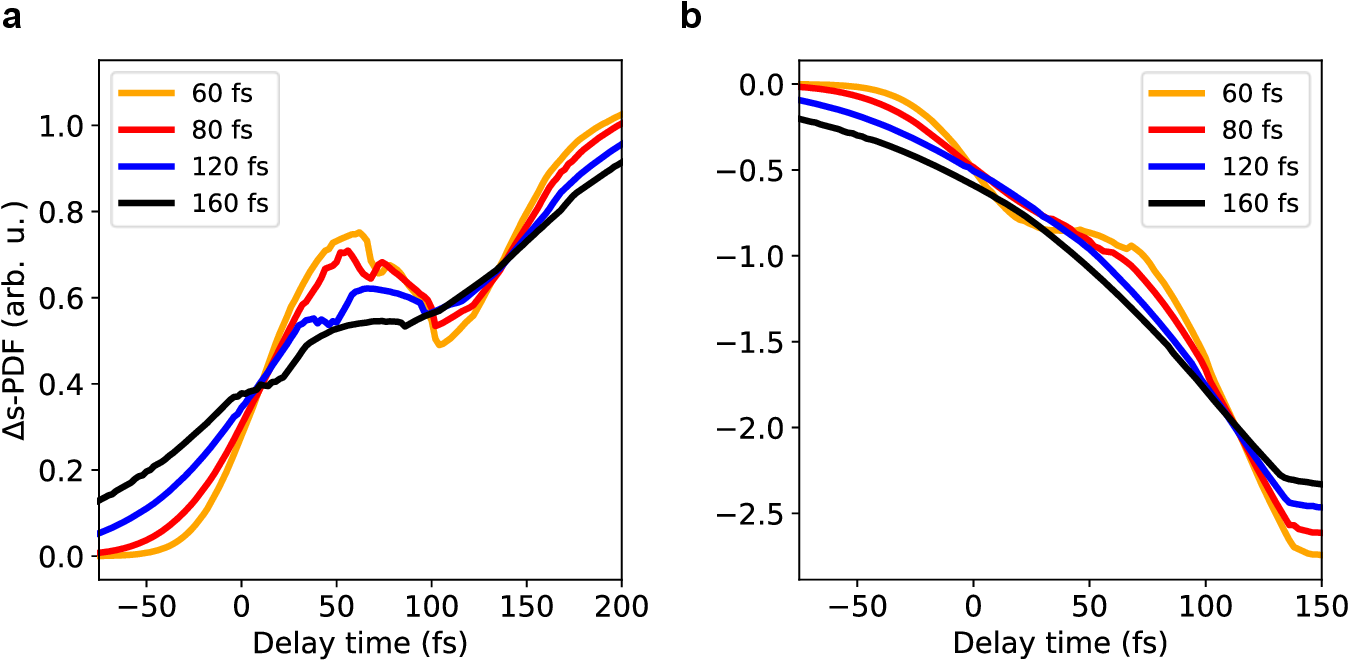} 
	\caption{\textbf{Simulated $\Delta$s-PDF after convolution with Gaussian kernels of various width.} \textbf{a} and \textbf{b}, Simulated intensity of $\Delta$$\text{s-PDF}$ at $\sim$1.95$\text{~\AA}$ and $\sim$2.30$\text{~\AA}$, respectively. The simulated results are convoluted with Gaussian kernels of 60~fs (orange), 80~fs (red), 120~fs (blue) and 160~fs (black) to illustrate the effect of temporal resolution in extracting the information of the CI dynamics. Source data are provided as a Source Data file.
    }
	\label{sup_temp} 
\end{figure}

\begin{figure} 
	\centering
	\includegraphics[width=0.8\textwidth]{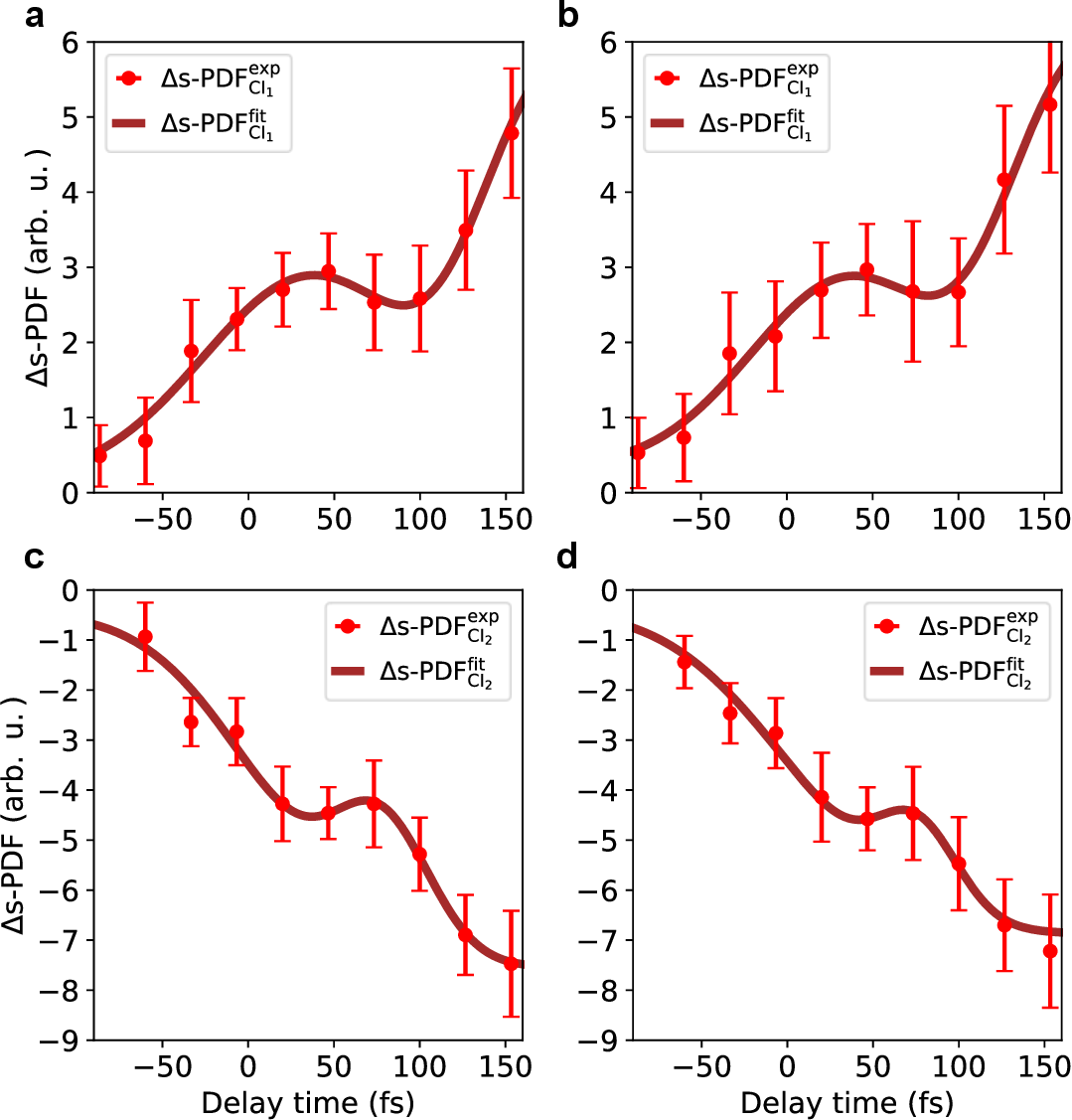} 
	\caption{\textbf{Structural dynamics during passage through the two CIs calculated by two different $s$-ranges.} \textbf{a} and \textbf{c}, Measured intensity of $\Delta$s-PDF at $\sim$1.95$\text{~\AA}$ and $\sim$2.30$\text{~\AA}$, respectively. The lower limit is taken as 1.5$\text{~\AA}^{-1}$. b and d, Measured intensity of $\Delta$s-PDF at $\sim$1.95$\text{~\AA}$ and $\sim$2.30$\text{~\AA}$, respectively. The lower limit is taken as 1.7$\text{~\AA}^{-1}$. Error bars represent a 68\% confidence interval calculated from bootstrap sampling. Source data are provided as a Source Data file.
    }
	\label{sup_srange} 
\end{figure}

\clearpage

\begin{table} 
	\centering
	\caption{\textbf{The energies of three lowest electronic states at S$_0$-min, CI$_1$, and CI$_2$.}
		The energy of the ground state minimum of CHD is set to zero.}
	\label{ext_table_MECI} 

	\begin{tabular}{cccc} 
		\\
		\hline
		Geometries & S$_0$~(eV) & S$_1$~(eV) & S$_2$~(eV)\\
		\hline
		S$_0$-min & 0 & 4.95 & 6.29\\
		CI$_1$ & 2.74 & 4.23 & 4.23\\
		CI$_2$ & 3.66 & 3.66 & 5.89\\
		\hline
	\end{tabular}
\end{table}

\begin{table} % Do not use \begin{table*}
	\centering
	% Captions go above tables
	\caption{\textbf{Characters of two low-lying electronic states.}
		The vertical excitation energies (VEEs), oscillator strength (f) and electronic characters of two low-lying electronic states at the XMS (3)-CASPT2 (6, 6)/def2-TZVP level at S$_0$-min of CHD.}
	\label{ext_table_estates} 

	\begin{tabular}{cccc} 
		\\
		\hline
		 & VEEs~(eV) & f & Dominant Contributions\\
		\hline
		S$_1$ & 4.95 & 0.16378 & HOMO$\to$LUMO (Single excitation)\\
		S$_2$ & 6.29 & 0.00197 & HOMO$\to$LUMO (Double excitation)\\
        & & & HOMO-1$\to$LUMO (Single excitation)\\
		\hline
	\end{tabular}
\end{table}

\end{document}